%% file: TOPQ_2014_17.tex
\pdfoutput=1
\newcommand*{\ATLASLATEXPATH}{}

\documentclass[cernpreprint,texlive=2016,txfonts,UKenglish]{\ATLASLATEXPATH atlasdoc}

\usepackage[block=space]{\ATLASLATEXPATH atlaspackage}
\usepackage{\ATLASLATEXPATH atlasbiblatex}
\usepackage{\ATLASLATEXPATH atlasphysics}
\newcommand{\Table}[1]{{Table~\ref{#1}}}
\newcommand{\Figure}[1]{{Figure~\ref{#1}}}
\newcommand{\Section}[1]{{Section~\ref{#1}}}

\newcommand{\Vlr}{V_{\text{L,R}}}
\newcommand{\Glr}{g_{\text{L,R}}} 
\newcommand{\ProjR}{P_{\text{R}}}
\newcommand{\ProjL}{P_{\text{L}}}
\newcommand{\ProjLR}{P_{\text{{L,R}}}}

\def\vl{\ensuremath{V_{\mathrm{L}}}}
\def\vr{\ensuremath{V_{\mathrm{R}}}}
\def\gr{\ensuremath{g_{\mathrm{R}}}}
\def\gl{\ensuremath{g_{\mathrm{L}}}}
\def\FL{\ensuremath{F_{\mathrm{L}}}}
\def\FR{\ensuremath{F_{\mathrm{R}}}}
\def\F0{\ensuremath{F_{0}}}
\def\AFB{\ensuremath{A_{\mathrm{FB}}}}
\def\AEC{\ensuremath{A_{\mathrm{EC}}}}

\input{TOPQ_2014_17-metadata}
\hypersetup{pdftitle={ATLAS document},pdfauthor={The ATLAS Collaboration}}

\begin{document}

\maketitle
\tableofcontents
\clearpage

\input{Introduction}

\input{Observables}
\input{Detector}

\input{Samples}

\input{Selections}

\input{Backgrounds}
\input{Yields}
\input{Distributions} 
\input{Unfolding}

\input{Uncertainties}
\FloatBarrier
\input{Results}

\FloatBarrier
\input{Conclusion}

\section*{Acknowledgements}
\input{Acknowledgements}

\clearpage
\printbibliography

\newpage
\input{atlas_authlist}

\end{document}

%% file: TOPQ_2014_17-metadata.tex
\AtlasTitle{Probing the $Wtb$ vertex structure in $t$-channel single-top-quark production and decay in $pp$ collisions at $\sqrt{\mathrm{s}}=8$~TeV with the ATLAS detector}
\author{The ATLAS Collaboration}

\PreprintIdNumber{CERN-EP-2017-011}

\AtlasJournal{JHEP}

\AtlasAbstract{To probe the $Wtb$ vertex structure, top-quark and $W$-boson 
polarisation observables are measured from $t$-channel single-top-quark events produced in 
proton--proton collisions at a centre-of-mass energy of 8~TeV. The dataset corresponds 
to an integrated luminosity of 20.2~fb$^{-1}$, recorded with the ATLAS detector at the LHC. 
Selected events contain one isolated electron or muon, large missing transverse momentum and 
exactly two jets, with one of them identified as likely to contain a $b$-hadron. Stringent 
selection requirements are applied to discriminate $t$-channel single-top-quark events from 
background. The polarisation observables are extracted from asymmetries in angular distributions 
measured with respect to spin quantisation axes appropriately chosen for the top quark and the 
$W$ boson. The asymmetry measurements are performed at parton level by correcting the observed angular 
distributions for detector effects and hadronisation after subtracting the background contributions. 
The measured top-quark and $W$-boson polarisation values are in agreement with the Standard 
Model predictions. Limits on the imaginary part of the anomalous coupling \gr\ are also set from 
model-independent measurements.}

%% file: Introduction.tex
\section{Introduction}
\label{sec:Introduction}

At hadron colliders, top quarks are predominantly produced in pairs (\ttbar) 
via the flavour-conserving strong interaction, but single top-quark production 
can occur via charged-current electroweak processes involving a $Wtb$ vertex. At 
leading order in QCD perturbation theory, three sub-processes contribute to single 
top-quark production: an exchange of a virtual $W$ boson either in the $t$-channel 
or in the $s$-channel, or the associated production of a top quark with an 
on-shell $W$ boson ($Wt$). The $t$-channel and $s$-channel processes do not
interfere at next-to-leading-order in QCD and are thus well defined
with that precision~\cite{Theory_sgtop}.

\begin{figure}[htbp]
\centering
\includegraphics[width= 0.30\textwidth]{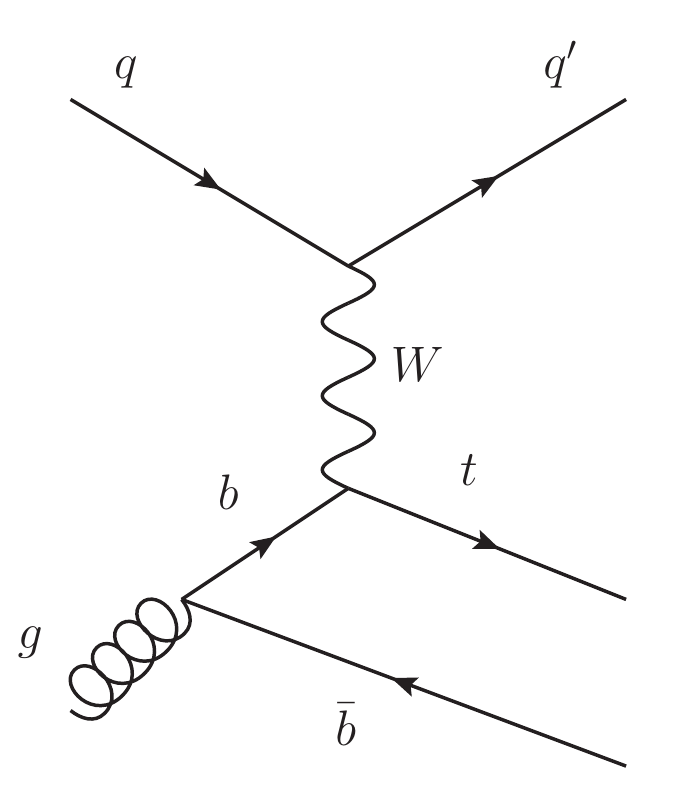}
\caption{Leading-order Feynman diagram for $t$-channel production of
single top quarks in $pp$ collisions. In the depicted four-flavour scheme (2$\rightarrow$3 process)
the initial $b$-quark arises from a gluon splitting into a $b\overline{b}$ pair.}
\label{fig:tchannel}
\end{figure}

In proton--proton ($pp$) collisions, the $t$-channel exchange, depicted in 
Figure~\ref{fig:tchannel}, is the dominant production process of single top quarks. 
The exchange of a space-like $W$ boson due to the interaction of a light quark with
a $b$-quark produces a top quark and a forward light-quark (called the spectator quark) 
in the final state. Furthermore, as a consequence of the vector minus axial-vector (V--A) 
form of the $Wtb$ vertex in the Standard Model, the produced top quarks are 
highly polarised, in particular along the direction of the spectator-quark 
momentum~\cite{Mahlon, Schwienhorst}. 

Within the Standard Model the top quark decays through the electroweak interaction 
into an on-shell $W$ boson and a $b$-quark, with a lifetime much shorter than 
the time scale necessary to depolarise the spin. The information on the top-quark 
spin can thus be obtained from its decay products. The produced real $W$ boson also 
possesses a polarisation (or helicity state), which can be extracted from angular 
distributions of its decay products through the measurement of spin-dependent 
observables~\cite{Theory_Wspin}.

Measuring the top-quark polarisation and the $W$-boson spin observables in $t$-channel 
single top-quark production provides a powerful probe for studying the $Wtb$ vertex 
in both top-quark production and decay. New physics effects resulting in corrections 
to the $Wtb$ vertex would affect the top-quark and $W$-boson polarisations. In the 
effective operator formalism the most general $Wtb$ Lagrangian can be written as~\cite{Theory_Wpolarization}:

\begin{equation}
{\cal L}_{Wtb} = - \frac{g}{\sqrt{2}}\,{\overline{b}}\gamma^\mu \left (\vl \ProjL + \vr \ProjR \right )tW^-_\mu 
- \frac{g}{\sqrt{2}}\,{\overline{b}}\,\frac{i\sigma^{\mu\nu}q_{\nu}}{\mW} \left (\gl \ProjL + \gr \ProjR \right)tW^-_\mu + \text{h.c.}
\label{eq:lagrangian}
\end{equation}

\noindent
In this expression $g$ is the weak coupling constant, $\mW$ and
$q_{\nu}$ are the mass and the four-momentum of the $W$ boson,
respectively, $\ProjLR \equiv (1\mp \gamma^5)/2$ are the left- and
right-handed projection operators, and $\sigma^{\mu\nu} =[\gamma^{\mu}, \gamma^{\nu}]/2$.  
The constants $\Vlr$ and $\Glr$ are the left- and right-handed 
vector and tensor couplings, respectively. In the Standard Model at 
tree level the coupling $\vl$ is the $V_{tb}$ element of the 
quark-mixing Cabibbo--Kobayashi--Maskawa (CKM) matrix that is close 
to one, while the anomalous couplings $\vr$ and $\Glr$ are all 
zero. Deviations from these values would provide hints of physics 
beyond the Standard Model, and complex values would imply that the 
top-quark decay has a CP-violating component~\cite{Theory_Wpolarization}. 
The imaginary part of $\gr$ (Im\,\gr) can be probed with the best precision 
in the $t$-channel production of single top quarks through the measurement 
of polarisation observables~\cite{Theory_Wpolarization}. Limits on Im\,\gr\ 
have been set at the LHC by the ATLAS Collaboration at a centre-of-mass energy 
of 7~TeV from the analysis of the double-differential angular decay rates 
of the produced $t$-channel single-top-quark events~\cite{ATLAS_Wtb_limits}.

The top-quark polarisation and the $W$-boson spin observables can
be extracted in an alternative way from the measurement of asymmetries 
in various angular distributions of the top-quark decay products~\cite{Theory_Wspin,Theory_Wpolarization}. 
Firstly, this article reports a determination of the top-quark polarisation 
as well as the $W$-boson spin observables extracted from the measured 
angular asymmetries. Such measurements serve as a consistency check 
with the Standard Model predictions. Secondly, limits on Im\,\gr\ are 
presented from the measurement of the so-called normal forward-backward 
asymmetry, which is predicted to have the highest sensitivity to 
Im\,\gr~\cite{Theory_Wpolarization}, and the asymmetry related to the 
top-quark polarisation. Here Standard Model values are assumed for all 
other couplings.

The measurements reported in this article use 20.2~fb$^{-1}$ of data collected 
at a centre-of-mass energy of 8~TeV with the ATLAS detector at the LHC. 
Stringent selection requirements are applied in order to separate signal from 
background. The $W$ boson from the top-quark decay is identified through 
its decay modes leading to a final state with an electron or a muon, and 
missing transverse momentum for the neutrino. The measurement at parton 
level of the asymmetries is performed by unfolding the 
observed angular distributions from detector and physics effects 
after subtracting the background contributions. For all reported results 
the electron and muon channels are merged, and the analysis is carried out 
independently of the lepton charge, in order to measure the polarisation 
observables associated with the combined production and decay of top 
quarks and top antiquarks.

%% file: Observables.tex
\section{Polarisation observables and asymmetries}
\label{sec:Observables}

The top-quark polarisation is determined from angular distributions of the decay products 
reconstructed in the top-quark rest frame, while the $W$-boson spin observables are 
determined from angular distributions of the charged lepton reconstructed in the $W$-boson 
rest frame.

In the top-quark rest frame, the angular distribution of any decay product $X$ of the top 
quark is given by

\begin{equation}
\frac{1}{\Gamma}\frac{\,\mathrm{d}\Gamma}{\,\mathrm{d}(\cos\theta_X)} = \frac{1}{2} \left(1 + \alpha_X P\cos\theta_X \right)\, ,
\label{eq:TopDecay}
\end{equation}

\noindent
where $\theta_X$ is the angle between the top-quark spin axis and the direction of motion of 
the chosen decay product in the top-quark rest frame, $\Gamma$ is the total decay width of the 
top quark, $\alpha_X$ is the spin analysing power associated with $X$, and $P$ is the top-quark 
degree of polarisation. The charged lepton is the most sensitive spin analyser; at 
next-to-leading-order (NLO) precision in QCD its spin analysing power is 
$\alpha_{\ell^{\pm}}$\,$=$\,$\pm0.998$~\cite{Brandenburg}. In the $t$-channel, single top quarks 
are produced with a large degree of polarisation in the direction of motion of the spectator 
quark~\cite{Mahlon_3,Schwienhorst}. This direction is used to define the top-quark spin axis 
in this measurement. The corresponding degrees of polarisation calculated at NLO in QCD are 
$0.91$ and $-0.86$ for top-quark and top-antiquark production, respectively~\cite{Schwienhorst}.

In the framework of a general formalism developed in Ref.~\cite{Theory_Wspin}, the spin-density 
matrix elements for the $W$-boson helicity components 0, $\pm$1, resulting from the 
decay of polarised top-quarks, can be parameterised in terms of expectation values of six 
independent spin observables: $\langle S_{1,2,3}\rangle$, $\langle T_{0}\rangle$ and 
$\langle A_{1,2}\rangle$. With ($\theta_{\ell}^*,\phi_{\ell}^*$) denoting the polar and azimuthal 
angles of the charged-lepton momentum in the $W$-boson rest frame, the fully differential decay 
width of a $W$ boson can be written as

\begin{eqnarray}
\frac{1}{\Gamma}\frac{\,\mathrm{d}\Gamma}{\,\mathrm{d}(\cos\theta_{\ell}^*)\mathrm{d}\phi_{\ell}^*}
&=& \frac{3}{8 \pi} \Bigg\{ \frac{2}{3} + \frac{1}{\sqrt{6}} \langle T_{0}\rangle \left(3 \cos^2\theta_{\ell}^* - 1 \right) + \langle S_{3}\rangle \cos\theta_{\ell}^* \nonumber \\
&+& \langle S_{1}\rangle \cos\phi_{\ell}^* \sin\theta_{\ell}^*\ + \langle S_{2}\rangle \sin\phi_{\ell}^* \sin\theta_{\ell}^* \nonumber \\
&-& \langle A_{1}\rangle \cos\phi_{\ell}^* \sin2\theta_{\ell}^*\ - \langle A_{2}\rangle \sin\phi_{\ell}^* \sin2\theta_{\ell}^* \Bigg\}\, .
\label{eq:WDecay}
\end{eqnarray}

\begin{figure}[!h!tpb]
\centering
\includegraphics[width=0.70\textwidth]{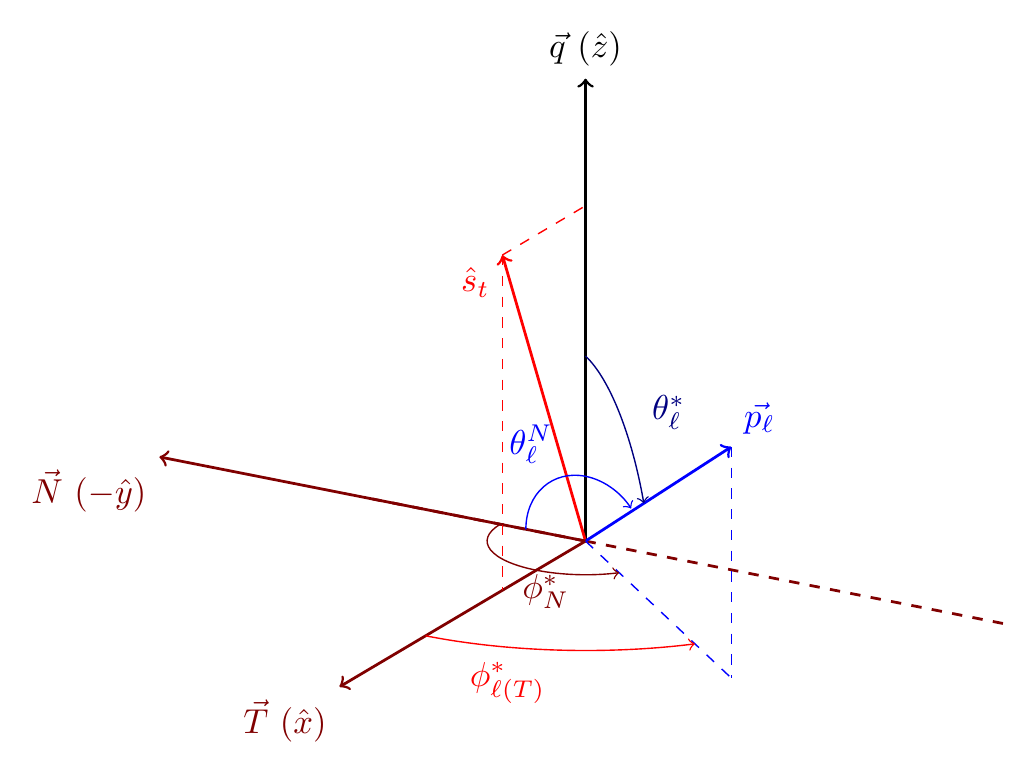}
\caption{Coordinate system and angles used to define the $W$-boson spin observables 
and their related angular asymmetries in the decay of polarised top quarks. The $W$-boson 
momentum ${\vec q}$ in the top-quark rest frame defines the $\hat{z}$-axis; the 
top-quark spin direction ${\hat{s}}_t$, taken along the spectator-quark momentum 
in the top-quark rest frame, defines the $\hat{x}$--$\hat{z}$ plane. The polar 
and azimuthal angles of the charged-lepton momentum ${\vec p}_{\ell}$ in the $W$-boson 
rest frame are labelled $\theta_{\ell}^*$ and $\phi_{\ell}^*$, respectively. The normal 
and transverse axes are defined relatively to ${\vec q}$ and ${\hat{s}}_t$ according 
to ${\vec N}={\hat{s}}_t\times {\vec q}$ and ${\vec T}={\vec q} \times {\vec N}$; they 
are along the $-\hat{y}$ and $\hat{x}$ axes of the coordinate system, respectively. The 
azimuthal angles $\phi_{N}^*$ and $\phi_{T}^*$ of the charged lepton in the $W$-boson 
rest frame are defined relatively to the ${\vec N}$ and ${\vec T}$ axes, respectively 
($\phi_{T}^*\equiv\phi_{\ell}^*$), while $\theta_{\ell}^N$ and $\theta_{\ell}^T$ (not 
shown in the figure) are the relative angles between ${\vec p}_{\ell}$ and the ${\vec N}$ 
and ${\vec T}$ axes, respectively.}
\label{fig:CoordinateSystem} 
\end{figure}

\noindent
In this formalism the $W$-boson spin axis is taken along the direction of the $W$-boson 
momentum in the top-quark rest frame, or equivalently along the direction opposite to the 
$b$-quark momentum in the $W$-boson rest frame. The coordinate system used and the various 
angles defined for the charged lepton in the $W$-boson rest frame are depicted in \Figure{fig:CoordinateSystem}. 

The angular distribution expressed in Equation~\eqref{eq:WDecay} implies an integration over all 
the possible directions of the top-quark spin relative to the $W$-boson spin axis. 
The top-quark polarisation is propagated to the spin observables $\langle S_{1,2}\rangle$ and 
$\langle A_{1,2}\rangle$, which depend in a proportional way on the value of $P$. The spin 
observables $\langle S_{3}\rangle$ and $\langle T_{0}\rangle$ do not depend on $P$, and are 
related to the $W$-boson helicity fractions \FR, \FL\ and \F0~\cite{Theory_Wspin}.

From the values of the helicity fractions predicted by the Standard Model at next-to-next-to-leading 
order (NNLO) in QCD assuming a top-quark mass of 172.5~GeV and a $b$-quark mass of 4.8~GeV~\cite{Theory_helicity}, 
one obtains $\langle S_{3}\rangle$\,$=-0.31$ and $\langle T_{0}\rangle$\,$=-0.43$. The uncertainties 
in these predictions due to the theoretical uncertainties in the helicity fractions are lower than 
0.01 for both $\langle S_{3}\rangle$ and $\langle T_{0}\rangle$. Combining the predicted degrees of 
polarisation $P_{t}=0.91$ and $P_{\bar t}=-0.86$ with the $t$-channel single-top cross-sections 
$\sigma_{t}=54.9$~pb and $\sigma_{\bar t}=29.7$~pb calculated at NLO in QCD for top-quark and 
top-antiquark production~\cite{Campbell_tchan}, the Standard Model predictions for $\langle S_{1,2}\rangle$ 
and $\langle A_{1,2}\rangle$ are: $\langle S_{1}\rangle$\,$=0.46$, $\langle A_{1}\rangle$\,$=0.23$ 
and $\langle S_{2}\rangle$\,$=$\,$\langle A_{2}\rangle$\,$=0$. These values are calculated at 
leading order (LO) in QCD from the expressions of the spin-density matrix elements given in Refs.~\cite{Theory_Wpolarization,Theory_Wspin}. 
The uncertainties in these predictions resulting from the uncertainties in the top-quark, $b$-quark 
and $W$-boson masses, and from higher-order effects~\cite{Theory_TopDecay}, are all smaller than 0.01. 
Measured values not equal to zero for the $\langle S_{2}\rangle$ 
and $\langle A_{2}\rangle$ spin observables would signal the presence of an imaginary coupling 
in the $Wtb$ vertex, since $\langle S_{2}\rangle$ and $\langle A_{2}\rangle$ are only sensitive 
to Im\,$\gr$~\cite{Theory_Wspin}.\footnote{Including one-loop QCD and electroweak corrections the 
prediction for \gr\ in the Standard Model is $(-7.17-1.23\mathrm{i})\times 10^{-3}$~\cite{ImGrPrediction}, 
leading to values of the order of $10^{-3}$ for the $\langle S_{2}\rangle$ and $\langle A_{2}\rangle$ 
spin observables.} However, $\langle S_{2}\rangle$ is twice as sensitive as $\langle A_{2}\rangle$ to Im\,$\gr$, 
making this observable more suitable for determining this coupling. The other four $W$-boson spin 
observables are mainly sensitive to Re\,$\gr$, with a poor sensitivity to Im\,$\gr$~\cite{Theory_Wpolarization,Theory_Wspin}.

\newcommand{\AsymmetryFootnote}{The asymmetries used in this article and in Ref.~\cite{Theory_Wpolarization}
are related to the ones defined in Refs.~\cite{Theory_Wspin,Theory_Zspin} through the equations 
$\AFB^{T}=\AFB^{x}$, $\AFB^{N}=-\AFB^{y}$, $\AFB^{T,\phi}=\AFB^{1}$, $\AFB^{N,\phi}=-\AFB^{2}$, $\AFB=\AFB^{z}$\,.}

The top-quark polarisation and the $W$-boson spin observables can be extracted from asymmetries 
derived by integrating the angular distributions expressed in Equations~\eqref{eq:TopDecay} and 
\eqref{eq:WDecay}. These asymmetries are based on single or combined angular observables. They 
are listed in \Table{tab:Asymmetries}, together with their associated angular observables and 
their relation to the polarisation observables.\footnote{\AsymmetryFootnote} The asymmetry values 
predicted by the Standard Model are also reported in the table.

Most of the polarisation observables are based on a forward-backward asymmetry, which is 
generically defined as a function of a given angular observable $\cos\theta$ according to

\begin{equation}
\AFB = \frac{N(\cos\theta>0)-N(\cos\theta<0)}{N(\cos\theta>0)+N(\cos\theta<0)}\, ,
\label{eq:AFB}
\end{equation}

\noindent
where $N$ is the number of events. One of the $W$-boson spin observables is determined 
from an asymmetry called edge-central and defined as follows

\begin{equation}
\AEC = \frac{N(|\cos\theta|>\frac{1}{2})-N(|\cos\theta|<\frac{1}{2})}{N(|\cos\theta|>\frac{1}{2})+N(|\cos\theta|<\frac{1}{2})}\, .
\label{eq:AEC}
\end{equation}

\begin{table}[h]
\begin{center}
\begin{tabular}{lccc}
\toprule
Asymmetry       & Angular observable                   & Polarisation observable                               & SM prediction \\
\midrule
$\AFB^{\ell}$   & $\cos\theta_{\ell}$                  & $\frac{1}{2}\alpha_{\ell} P$                          & 0.45    \\[0.14cm]
$\AFB^{tW}$     & $\cos\theta_W \cos\theta_{\ell}^*$   & $\frac{3}{8} P \left( \FR + \FL \right)$              & 0.10    \\[0.18cm]
$\AFB$          & $\cos\theta_{\ell}^*$                & $\frac{3}{4} \langle S_{3}\rangle = \frac{3}{4} \left( \FR - \FL \right)$                  & $-$0.23 \\[0.10cm]
$\AEC$          & $\cos\theta_{\ell}^*$                & $\frac{3}{8} \sqrt{\frac{3}{2}} \langle T_{0}\rangle = \frac{3}{16}\left(1 - 3\F0 \right)$ & $-$0.20 \\[0.14cm]
$\AFB^{T}$      & $\cos\theta_{\ell}^T$                & $\frac{3}{4} \langle S_{1}\rangle$                    & 0.34    \\[0.14cm] 
$\AFB^{N}$      & $\cos\theta_{\ell}^N$                & $-\frac{3}{4} \langle S_{2}\rangle$                   & 0       \\[0.14cm]        
$\AFB^{T,\phi}$ & $\cos\theta_{\ell}^* \cos\phi_{T}^*$ & $-\frac{2}{\pi} \langle A_{1}\rangle$                 & $-$0.14 \\[0.14cm]
$\AFB^{N,\phi}$ & $\cos\theta_{\ell}^* \cos\phi_{N}^*$ & $\frac{2}{\pi} \langle A_{2}\rangle$                  & 0       \\[0.14cm]
\bottomrule
\end{tabular}
\caption{Asymmetries with their associated angular observables and their relation to the top-quark 
polarisation and $W$-boson spin observables. The values predicted by the Standard Model are 
also given. They are calculated using the predictions at NLO in QCD for $P$ and $\alpha_{\ell}$, 
the predictions at NNLO for the helicity fractions, and the predictions at LO for 
$\langle S_{1,2}\rangle$ and $\langle A_{1,2}\rangle$. The uncertainties in these values 
are all lower than 0.01. They are estimated from the uncertainties in the top-quark, $b$-quark 
and $W$-boson masses, added in quadrature, including the uncertainty in $\alpha_\mathrm{s}$ and an 
estimate of the higher-order effects for the asymmetries related to the $W$-boson spin observables.}
\label{tab:Asymmetries}
\end{center}
\end{table}

The product $\alpha_{\ell} P$ is extracted from the forward-backward asymmetry $\AFB^{\ell}$ 
of the $\cos\theta_{\ell}$ angular distribution, where $\theta_{\ell}$ is the angle between 
the lepton momentum in the top-quark rest frame and the top-quark spin axis. The measurement 
of $P$ can also be performed from the forward-backward asymmetry $\AFB^{tW}$ defined with 
respect to the combined angular observable $\cos\theta_W \cos\theta_{\ell}^*$~\cite{Theory_TopSpin_1}, 
where $\theta_W$ is the angle between the $W$-boson momentum in the top-quark rest frame 
and the top-quark spin axis. This asymmetry is proportional to the product of $P$ and the sum 
of two $W$-boson helicity fractions, as reported in \Table{tab:Asymmetries}. The $W$-boson 
spin observables $\langle S_{3}\rangle$ and $\langle T_{0}\rangle$ are derived from the 
forward-backward asymmetry $\AFB$ and from the edge-central asymmetry $\AEC$ of the 
$\cos\theta_{\ell}^*$ angular distribution, respectively. Using the definition~\cite{Theory_Wpolarization} 
of the normal axis ${\vec N}={\vec s}_t \times {\vec q}$ and transverse axis ${\vec T}={\vec q} \times {\vec N}$, 
as illustrated in \Figure{fig:CoordinateSystem}, $\langle S_{1}\rangle$ and $\langle S_{2}\rangle$ 
are determined from the forward-backward asymmetries $\AFB^{T}$ and $\AFB^{N}$ in the angular 
observables $\cos\theta_{\ell}^T$ and $\cos\theta_{\ell}^N$, respectively. The $\langle A_{1}\rangle$ 
and $\langle A_{2}\rangle$ spin observables are determined from the forward-backward 
asymmetries $\AFB^{T,\phi}$ and $\AFB^{N,\phi}$ based on the combination of $\cos\theta_{\ell}^*$ 
with the cosine of the azimuthal angles $\phi_{T}^*$ and $\phi_{N}^*$ defined relatively 
to ${\vec T}$ and ${\vec N}$, respectively.

Limits on Im\,\gr\ can be extracted from the measurement of the $\AFB^{N}$ asymmetry, which 
has the highest sensitivity to this coupling. For small Im\,\gr\ values, taking $\vl=1$ and 
$\vr=\gl=0$, a linear dependence on Im\,\gr\ is obtained for this asymmetry: 
$\AFB^{N}=0.64\,P\,\mathrm{Im}\,\gr$~\cite{Theory_Wpolarization}. In this relation the 
weak dependence of $P$ on Im\,\gr, which is of quadratic form, is not included.
As $\AFB^{N}$ depends on $P$, the measured value of the $\AFB^{\ell}$ asymmetry is required 
to constrain $P$ for the limit computation. The quadratic variation of $P$ and $\alpha_{\ell}$ 
as a function of Im\,\gr~\cite{Theory_Wpolarization,Theory_TopSpin_2} is taken into account 
when setting the limits through the procedure explained in \Section{sec:Results}. The $\AFB^{\ell}$ 
asymmetry is chosen to constrain $P$ because it is measured independently of Im\,\gr; this 
is discussed in \Section{sec:Unfolding}.

%% file: Detector.tex
\section{The ATLAS detector}
\label{sec:Detector}

\newcommand{\AtlasCoordFootnote}{ATLAS uses a right-handed coordinate system with its origin 
at the nominal interaction point in the centre of the detector and the $z$-axis along the beam 
pipe. The $x$-axis points from the interaction point to the centre of the LHC ring, and the 
$y$-axis points upwards. Cylindrical coordinates $(r,\phi)$ are used in the transverse plane, 
$\phi$ being the azimuthal angle around the $z$-axis. The pseudorapidity is defined in terms 
of the polar angle $\theta$ as $\eta = -\ln \tan(\theta/2)$.}

The ATLAS detector~\cite{ATLAS_detector} is a multi-purpose particle detector with a forward-backward 
symmetric, cylindrical geometry and a near $4\pi$ coverage in solid angle around the collision 
point.\footnote{\AtlasCoordFootnote} It consists of an inner tracking detector surrounded by a thin 
superconducting solenoid, electromagnetic and hadronic calorimeters, and a muon spectrometer. The 
inner detector is immersed in a \SI{2}{\tesla} axial magnetic field, and provides charged-particle 
tracking in the pseudorapidity range $|\eta|<2.5$. It contains a high-granularity silicon pixel detector, 
a silicon microstrip tracker, and a straw-tube transition radiation tracker. Lead/liquid-argon 
sampling calorimeters provide electromagnetic energy measurements with high granularity in the 
pseudorapidity ranges $|\eta|<1.5$ (barrel region) and $1.4<|\eta|<3.2$ (endcap region). Hadronic 
energy measurements are provided by steel/scintillator-tile calorimeters in the central pseudorapidity 
range $|\eta|<1.7$ and by copper/liquid-argon calorimeters in the endcap region $1.5<|\eta|<3.2$.
The forward region is instrumented with liquid-argon calorimeters for electromagnetic and 
hadronic energy measurements, extending the coverage to $|\eta|=4.9$. The muon spectrometer surrounds 
the calorimeters and incorporates three large air-core toroid superconducting magnets with eight 
coils each. It includes separate trigger detectors and high-precision tracking chambers, providing 
muon momentum measurement for $|\eta|<2.7$ and muon triggering up to $|\eta|=2.4$.

A three-level trigger system is used to select interesting events~\cite{ATLAS_trigger}. The first-level 
trigger is hardware-based and uses a subset of the detector information to reduce the accepted event 
rate to less than \SI{75}{\kHz}. The second and third levels are software-based and together reduce 
the event rate to about \SI{400}{\Hz}.

%% file: Samples.tex
\section{Data and simulation samples}
\label{sec:Samples}

The analysis is performed using $pp$ collision data collected in 2012 by the ATLAS detector
at a centre-of-mass energy of $\SI{8}{\TeV}$. The events are required to pass single-electron 
or single-muon triggers~\cite{ATLAS_trigger,ATLAS_trigger_muon}, resulting, after detector and 
data-quality requirements, in a data sample corresponding to an integrated luminosity of 
$\SI{20.2}{\per\fb}$. The electron and muon triggers impose a threshold of $\SI{24}{\GeV}$ on 
the transverse momentum (\pT), along with isolation requirements. To recover efficiency for
higher-\pT\ leptons, the isolated lepton triggers are complemented by triggers without isolation 
requirements, but with a threshold raised to $\SI{60}{\GeV}$ for electrons and to $\SI{36}{\GeV}$ 
for muons.

Samples of signal and background events are simulated using various Monte Carlo generators.
The generated events are passed through a simulation of the ATLAS detector~\cite{ATLAS_FullSim} 
based on the \textsc{Geant4} framework~\cite{GEANT4}. For some samples a faster simulation 
(ATLFAST-II~\cite{ATLAS_FastSim}), making use of a parameterised response of the electromagnetic 
and hadronic calorimeters, is performed instead. Minimum-bias events simulated with 
the \textsc{Pythia}~(8.1)~\cite{Pythia8} generator are overlaid to model the pile-up effects 
from additional $pp$ collisions in the same and nearby bunch crossings. All simulated events are 
then processed using the same reconstruction and analysis chain as for data events.
  
Signal $t$-channel single-top-quark events are generated with the NLO \textsc{Powheg-Box}~(r2556)~\cite{Powheg_1,Powheg_2,Powheg_3} 
generator, which uses the four-flavour scheme (Figure~\ref{fig:tchannel}) for the matrix-element calculations~\cite{Frederix:2012}. 
Events are generated with the CT10f4~\cite{PDF_Lai} parton distribution functions (PDFs), and the 
renormalisation and factorisation scales are set to $\mu_{\text{R}}^2=\mu_{\text{F}}^2=16\left(m^2_b+p^2_{\text{T},b}\right)$,
where $m_{b}$ is the mass of the $b$-quark and $p_{\text{T},b}$ is the transverse momentum of the $b$-quark 
from the initial gluon splitting (called the spectator $b$-quark)~\cite{Frederix:2012}. Additional 
$t$-channel samples are produced with the LO \textsc{Protos}~(2.2)~\cite{Protos} 
generator using the CTEQ6L1 PDFs~\cite{PDF_CTEQ6L1}. \textsc{Protos} events are generated using 
the four-flavour scheme, as well, and anomalous couplings are enabled in both the production and the decay 
vertices, varying Re\,$\vl$ and Im\,$\gr$ simultaneously to keep the top-quark width invariant. 
The factorisation scale is set to $\mu_{\text{F}}^2=-p^2_W$ for the light quark, where $p_W$ is 
the four-momentum of the exchanged $W$ boson, and to $\mu_{\text{F}}^2=m^2_b+p^2_{\text{T},b}$ 
for the gluon. Eight \textsc{Protos} samples generated with Im\,$\gr$ in the range [$-$0.144, 0.144] 
and Re\,$\vl$ in the range [0.982, 1] are used, including the Standard Model configuration 
Im\,$\gr=0$ and Re\,$\vl=1$. These \textsc{Protos} samples are used to compute the parton-level 
unfolding corrections and to check the reliability of the unfolding method, while the 
\textsc{Powheg-Box} sample is used to determine the expected event yields and template 
distributions.

Samples of \ttbar~\cite{Powheg_ttbar}, $s$-channel single-top-quark and $Wt$~\cite{Powheg_Wt} 
background events are produced using the \textsc{Powheg-Box}~(r2819, r3026) generator with the 
CT10 PDFs. To generate the \ttbar\ sample, the model parameter $h_{\text{damp}}$, which effectively 
regulates the high-\pT\ gluon radiation, is set to the top-quark mass $m_{t}$~\cite{ATLAS_hdamp}. 

For the above samples, parton showering, hadronisation and the underlying event are simulated 
with \textsc{Pythia}~(6.426)~\cite{Pythia} using parameter values set to the Perugia 2011C 
tune~\cite{PerugiaTune}, and the CTEQ6L1 PDFs.

To study the modelling uncertainties of all processes involving top quarks, either alternative 
generators or parameter variations in the \textsc{Powheg-Box} and \textsc{Pythia} settings are 
used. For the estimation of the uncertainty in the $t$-channel matrix-element calculation, a 
sample is produced using the \textsc{MadGraph5}\_a\textsc{MC@NLO}~(2.0)~\cite{aMCatNLO}
generator, interfaced to \textsc{Herwig}~(6.52)~\cite{Herwig_1,Herwig_2} for parton showering 
and to \textsc{Jimmy}~(4.31)~\cite{Jimmy} for the underlying-event modelling with the ATLAS AUET2 
tuned parameter settings~\cite{AUETTune} and the CT10f4 PDFs. The events are generated using 
the four-flavour scheme. For the \ttbar, $s$-channel and $Wt$ processes, alternative samples 
are produced using the \textsc{MC@NLO}~(4.03)~\cite{MCatNLO_1,MCatNLO_2,MCatNLO_3,MCatNLO_4} 
generator interfaced to \textsc{Herwig}~(6.52) for parton showering and \textsc{Jimmy}~(4.31) 
for the underlying-event modelling with the ATLAS AUET2 tune and the CT10 PDFs. To specifically 
study the impact of the parton-shower modelling, a $t$-channel sample and a $Wt$ sample both
generated with \textsc{Powheg-Box} and coupled to \textsc{Herwig}~(6.52) and \textsc{Jimmy}~(4.31) 
with the AUET2 tune are used. For the \ttbar\ process, samples generated using \textsc{Powheg-Box} 
with the CT10 PDFs, interfaced to \textsc{Herwig}~(6.52) with the AUET2 tune or to \textsc{Pythia}~(6.426) 
with the AUET2B tune, are used. Effects of varying the amount of radiation are studied by changing 
the hard-process and parton-shower scales simultaneously in the \textsc{Powheg-Box} and \textsc{Pythia}~(6.426, 6.427) 
simulations. In the single-top-quark samples the factorisation and renormalisation scales are increased 
or decreased by a factor of two or one-half, respectively, in combination with the Perugia 2012 radLo 
and radHi tunes~\cite{PerugiaTune}. In the \ttbar\ samples, $h_{\text{damp}}$ is set to $m_{t}$
or $2m_{t}$ in combination with the radLo and radHi parameterisations, respectively. 

All top-quark processes are simulated assuming a top-quark mass of 172.5~GeV, and the top-quark decay is 
assumed to proceed exclusively through \textit{t}$\to$\textit{Wb}. The baseline \textsc{Powheg-Box} samples 
are passed through the fully \textsc{Geant4}-based simulation of the ATLAS detector, while the \textsc{Protos} samples 
and all samples used in studies of modelling uncertainties are processed through the ATLFAST-II simulation.
    
Vector-boson production in association with jets is simulated using the multileg LO \textsc{Sherpa}~(1.4.1)~\cite{Sherpa} 
generator with its own parameter tune and the CT10 PDFs. \textsc{Sherpa} is used not only to generate the hard 
process, but also for the parton shower and the modelling of the underlying event. $W$+jets and $Z$+jets events 
with up to four additional partons are generated. The CKKW method~\cite{Hoeche:2009} is used to remove overlaps 
between the partonic configurations generated by the matrix element and by the parton showering. Diboson samples 
of $WW$, $WZ$ and $ZZ$ events are also produced, using the \textsc{Sherpa}~(1.4.1) generator with the CT10 PDFs. 
All the generated \textsc{Sherpa} single-boson and diboson events are passed through the ATLFAST-II simulation 
of the detector.

%% file: Selections.tex
\section{Event reconstruction and selection}
\label{sec:Selections}

The analysis considers only $W$-boson decay modes to an electron or a muon. Events in 
which the $W$ boson decays to a $\tau$ lepton are thus included if the $\tau$ lepton 
subsequently decays to an electron or a muon. 

The signal event candidates are selected by requiring a single isolated electron or muon, 
significant missing transverse momentum, and exactly two jets with one of them identified 
as likely to contain a $b$-hadron ($b$-tagged jet). In fact, the presence of a third jet is 
not required in the event selection. Indeed, the additional jet resulting from the 
spectator $b$-quark originating from the gluon splitting as shown in \Figure{fig:tchannel} 
is expected to have a softer \pT spectrum and a broader $|\eta|$ distribution than 
the $b$-tagged jet produced in the top-quark decay, and, therefore, is in general not detected.

Electron candidates are reconstructed from isolated energy deposits in the electromagnetic 
calorimeter which are associated with inner-detector tracks fulfilling strict quality 
requirements~\cite{ATLAS_electrons}. They are required to satisfy $\pT>\SI{25}{\GeV}$ and 
$|\eta|<2.47$, excluding the barrel--endcap transition region, corresponding to $1.37<|\eta|<1.52$. 
Muon candidates are reconstructed using combined tracking information from the inner detector
and the muon spectrometer~\cite{ATLAS_muons}. They are required to have $\pT>\SI{25}{\GeV}$ 
and $|\eta|<2.5$. The electron and muon candidates must fulfil additional isolation 
requirements, as described in Ref.~\cite{ATLAS_topreco}, in order to reduce contributions 
from misidentified jets, non-prompt leptons from the decay of heavy-flavour quarks and 
electrons from photon conversions.

\newcommand{\VertexFootnote}{A primary-vertex candidate is defined as a reconstructed 
vertex with at least five associated tracks with $\pT>400$~MeV. The primary vertex 
associated with the hard-scattering collision is the candidate with the largest sum 
of the squared \pT of the associated tracks.}

Jets are reconstructed using the anti-$k_{t}$ algorithm~\cite{Antikt} with a radius 
parameter of 0.4, from topological clusters~\cite{ATLAS_clusters}, calibrated with 
a local cluster weighting method~\cite{ATLAS_jets_1}. Jets are calibrated using an 
energy- and $\eta$-dependent simulation-based scheme, with in situ corrections based 
on data. The jet energy is further corrected for the effect of multiple $pp$ interactions. 
To reject jets from pile-up events, a so-called jet-vertex-fraction criterion~\cite{ATLAS_jvf} 
is applied to the jets with $\pT<\SI{50}{\GeV}$ and $|\eta|<2.4$: at least 50\% of the 
scalar sum of the \pT\ of the tracks associated with a 
jet is required to be from tracks compatible with the primary vertex.\footnote{\VertexFootnote}
Only events containing two reconstructed jets with $\pT>\SI{30}{\GeV}$ are selected. 
In addition, one of them must be $b$-tagged with $|\eta|<2.5$, while the 
second jet is required to be untagged and to have $|\eta|<4.5$. The $b$-tagging is 
performed using a neural network which combines three different algorithms exploiting 
the properties of a $b$-hadron decay in a jet~\cite{ATLAS_btag_1}. The $b$-tagging 
algorithm is optimised to improve the rejection of $c$-quark jets, since $W$-boson 
production in association with $c$-quarks is a major background for the selected final 
state. The requirement applied to the neural-network discriminant corresponds to a 
$b$-tagging efficiency of 50\%, and mistagging rates of 3.9\% and 0.07\%
for $c$-quark jets and light-flavour jets, respectively, as predicted in simulated 
\ttbar events~\cite{ATLAS_btag_2,ATLAS_btag_3}.

The missing transverse momentum, with magnitude \MET, is reconstructed from the vector 
sum of energy deposits in the calorimeter projected onto the transverse plane~\cite{ATLAS_MET}.
All cluster energies are corrected using the local cluster weighting method. Clusters 
associated with high-\pT jets and electrons are further calibrated using their respective 
energy corrections. Contributions from the \pT of the selected muons are also included 
in the calculation. 

\newcommand{\MassFootnote}{The transverse mass of the lepton--\MET\ system is defined as
$m_{\mathrm{T}}(\ell, \MET) = \sqrt{2 \pT(\ell) \MET \left(1-\cos \Delta \phi (\ell,\MET) \right)}$,
where $\Delta \phi(\ell,\MET)$ is the difference in azimuthal angle between the lepton 
transverse momentum and the missing transverse momentum.}

Events are required to contain at least one good primary-vertex candidate, and no jets failing 
to satisfy reconstruction quality criteria. The magnitude of the missing transverse momentum 
is required to be larger than $\SI{30}{\GeV}$. In addition, the transverse mass of the 
lepton--\MET system must be greater than $\SI{50}{\GeV}$ in order to reduce the multijet 
background contribution.~\footnote{\MassFootnote} Further reduction of this background is achieved 
by imposing an additional requirement on events where the lepton and the leading jet in \pT have 
opposite directions in the transverse plane~\cite{ATLAS_tchan}. To reduce the \ttbar\ dilepton 
background, events containing an additional lepton, identified with less stringent 
criteria (referred to as a loose lepton) and with a \pT\ threshold lowered to $10$~GeV, 
are rejected.

The lepton and neutrino four-momenta are used to reconstruct the $W$ boson. Since the neutrino 
escapes undetected, the $x$- and $y$-components of the missing transverse momentum are assumed 
to correspond to the transverse momentum of the neutrino. The unmeasured longitudinal component 
of the neutrino momentum is computed by imposing a $W$-boson mass constraint on the lepton--neutrino 
system. If there are two real solutions, the solution giving the smallest magnitude of the 
longitudinal neutrino momentum is taken. If there are complex solutions, the magnitude of the 
measured missing transverse momentum is rescaled in order to obtain a physical solution~\cite{ATLAS_Wtb_limits}. 
The top-quark candidate is reconstructed by combining the four-momenta of the reconstructed $W$ 
boson and the $b$-tagged jet.

Additional requirements, defining the signal region, are finally applied to the preselected events:
   
\begin{itemize}
 
\item
The pseudorapidity of the untagged jet must satisfy $|\eta|>2.0$, since the spectator quark tends 
to be produced in the forward direction in the $t$-channel process.
\item
The separation in $\eta$ between the untagged jet and the $b$-tagged jet must be larger than 1.5,
to reduce the contribution from \ttbar\ background events.
\item
The mass of the reconstructed top quark is required to be between 130~GeV and 200~GeV, to reject 
background events from processes not involving top quarks.
\item
The scalar sum ($H_\mathrm{T}$) of the \pT\ of the lepton, the \pT\ of the jets and \MET must be larger than 195~GeV, to
further reduce the number of background events, in particular the $W$+jets contribution.
       
\end{itemize}

\begin{figure}[t]
\captionsetup[subfloat]{farskip=2pt,captionskip=1pt}
\centering
\subfloat[]
{\includegraphics[width=0.48\textwidth]{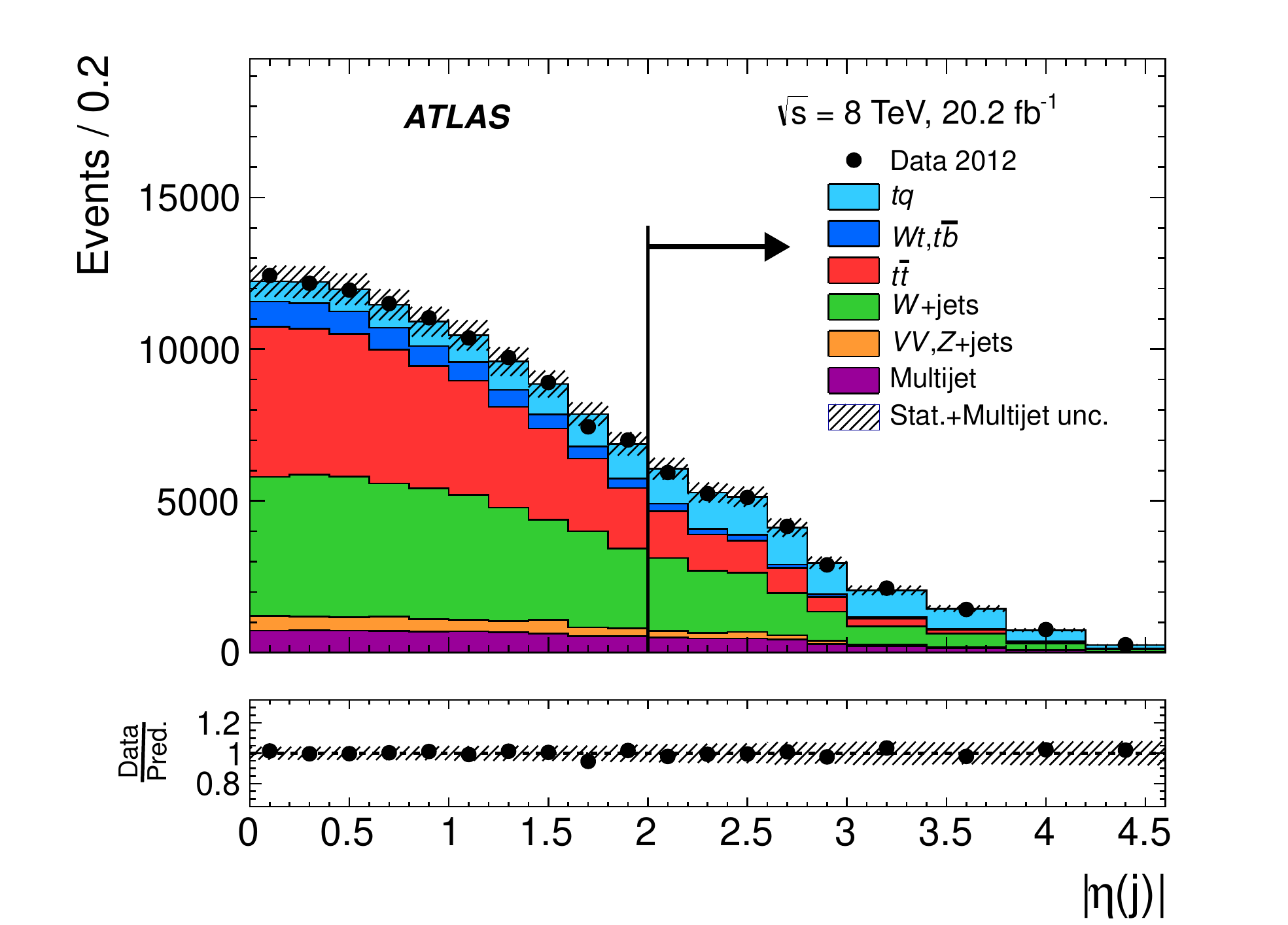}
\label{fig:preselection_abs_ujet_eta}}
\subfloat[]
{\includegraphics[width=0.48\textwidth]{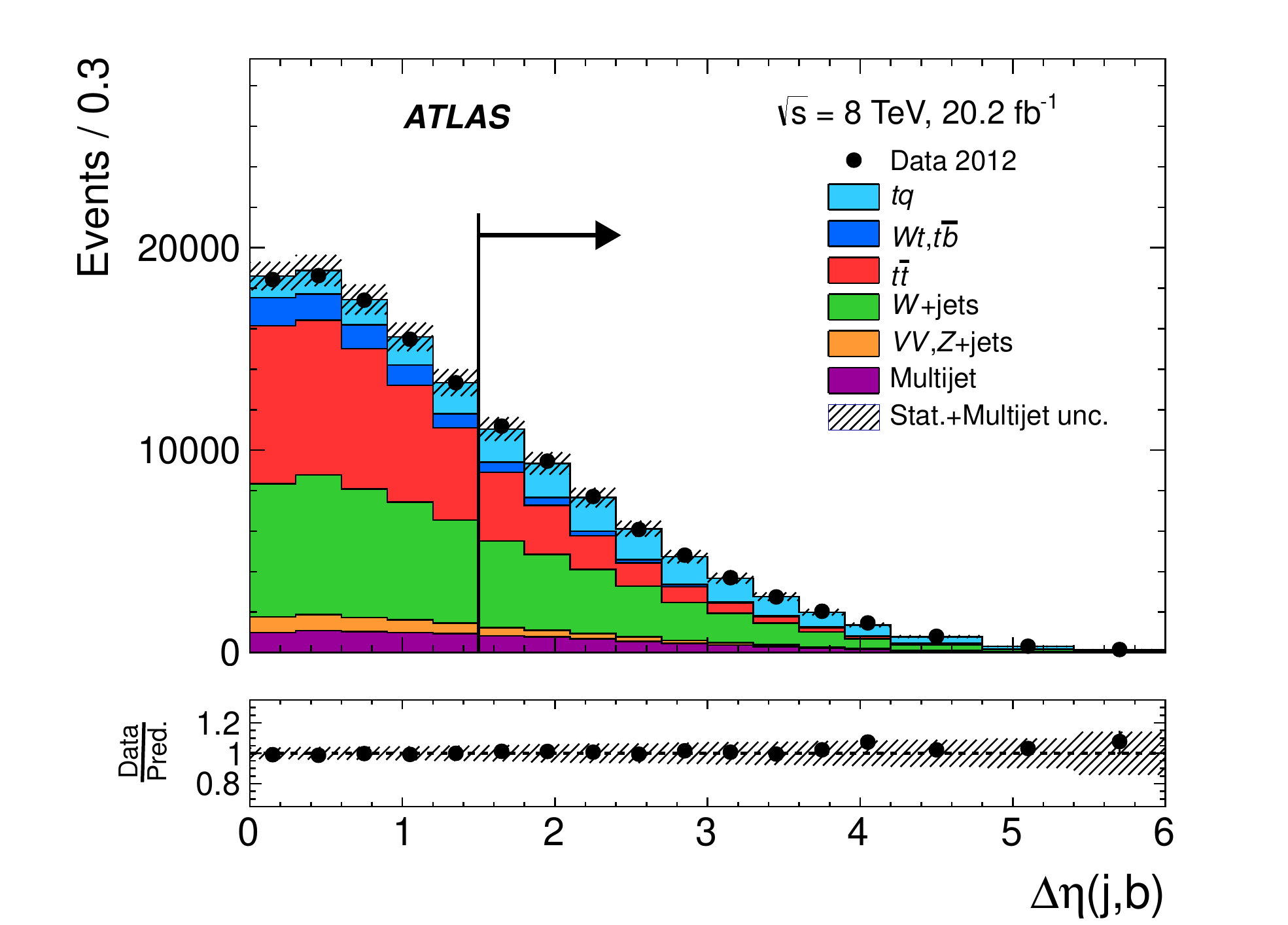}
\label{fig:preselection_dEta_bjet_ujet}}
\hfill
\subfloat[]
{\includegraphics[width=0.48\textwidth]{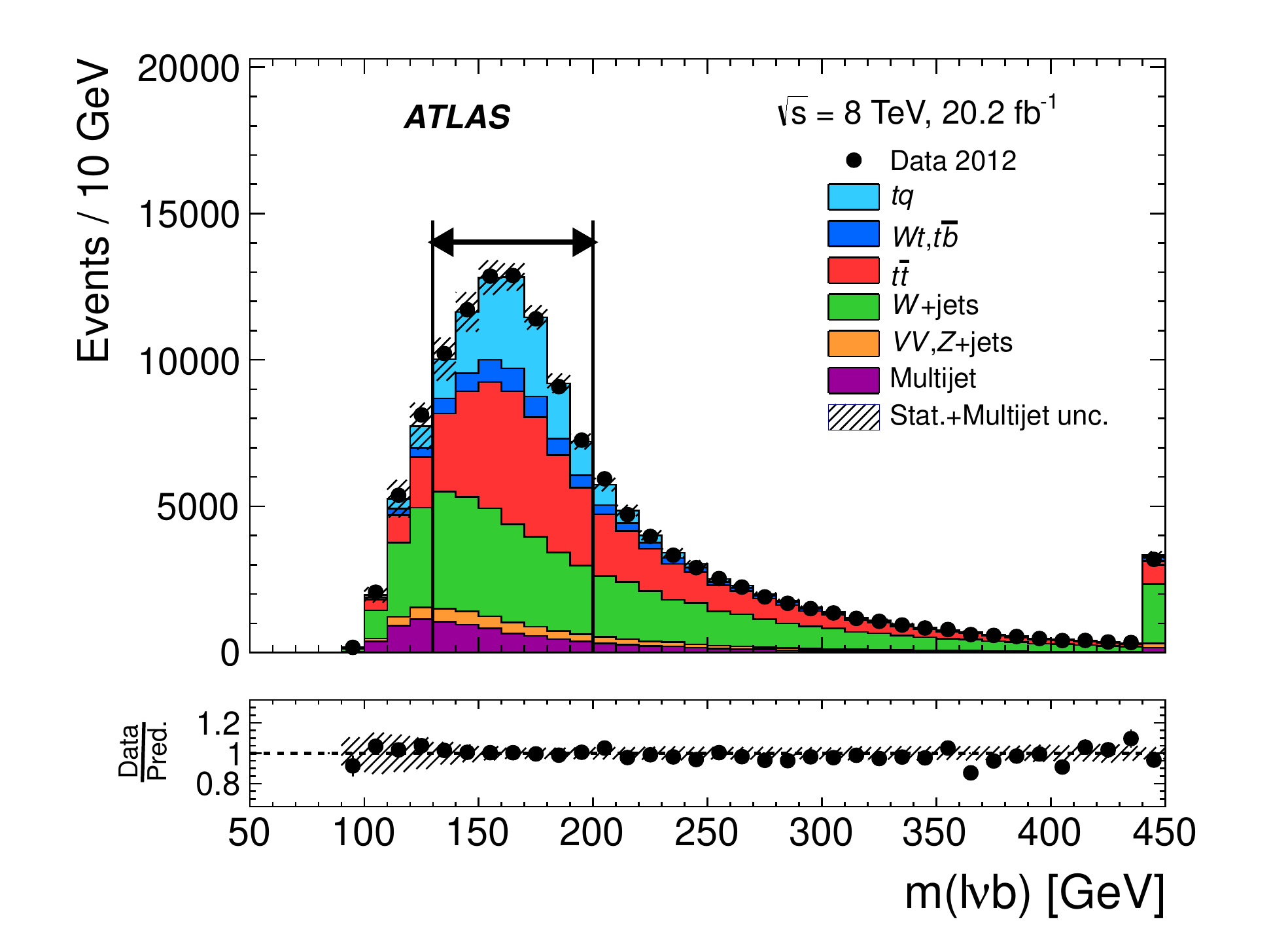}
\label{fig:preselection_top_m}}
\subfloat[]
{\includegraphics[width=0.48\textwidth]{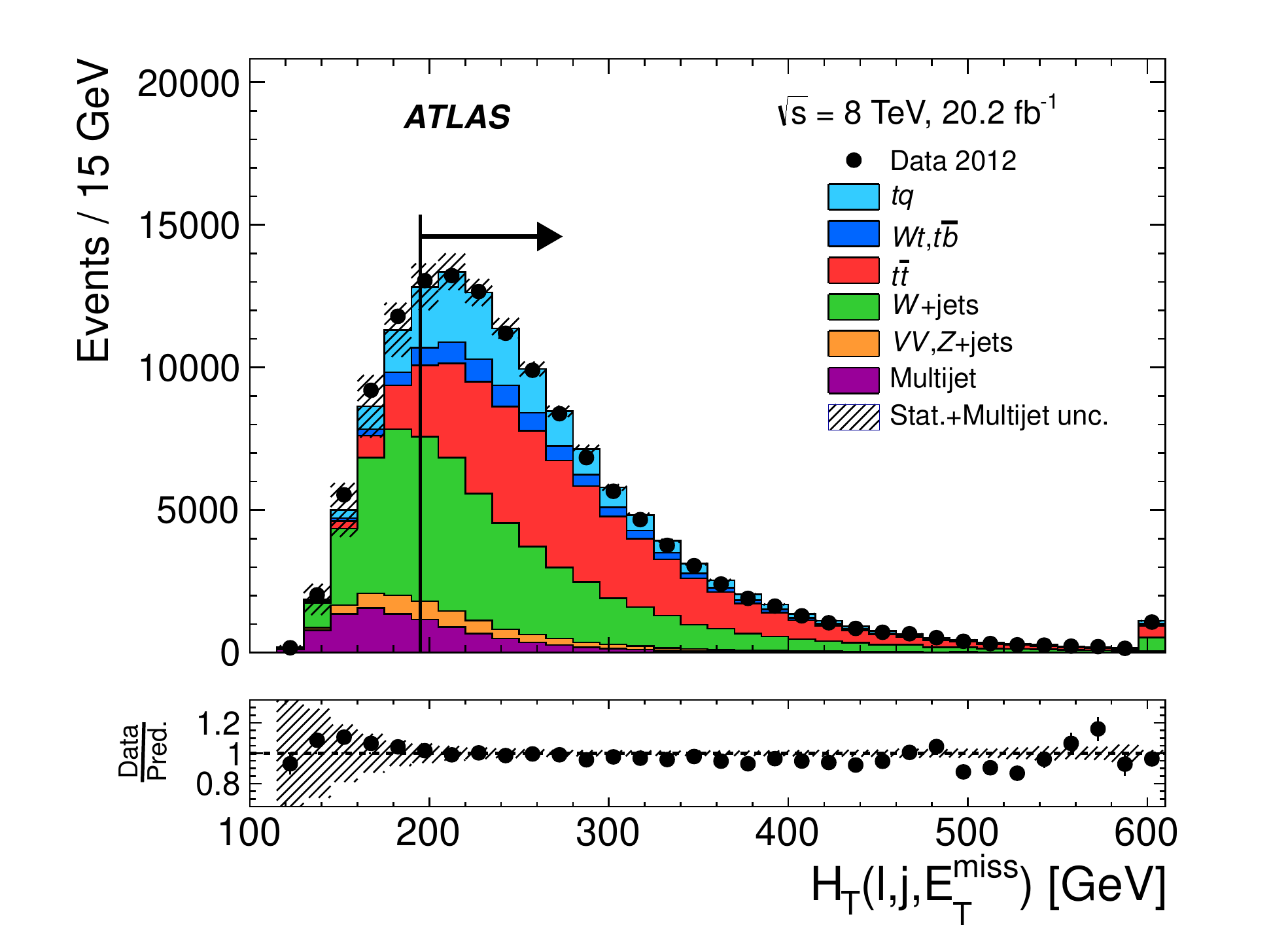}
\label{fig:preselection_Ht_event}}
\caption
{Distributions of the selection variables in the preselected signal region:
\protect\subref{fig:ttbar_abs_ujet_eta} $|\eta|$ of the untagged jet,
\protect\subref{fig:ttbar_dEta_bjet_ujet} separation in $\eta$ between the untagged and $b$-tagged jets,
\protect\subref{fig:ttbar_top_m} reconstructed top-quark mass, and
\protect\subref{fig:ttbar_Ht_event} scalar sum of the \pT\ of the lepton, the \pT\ of the jets and \MET.
The observed distributions are compared to the predicted signal and background distributions, 
normalised to the results of the maximum-likelihood fit. The labels $tq$ and $t\bar{b}$ refer 
to the $t$-channel and $s$-channel single-top-quark processes, respectively, and $VV$ to diboson 
production. The vertical lines and the arrows define the signal region. The uncertainty bands 
include the statistical post-fit uncertainty, the uncertainty due to the limited size of the simulation 
samples and the uncertainty in the normalisation of the multijet background, added in quadrature. The last 
bin of the histograms includes overflows. The lower panels show the ratio of data to prediction.}
\label{fig:selections}
\end{figure}

Figure~\ref{fig:selections} shows the distributions of the four variables relevant for these
requirements, comparing data to the predicted signal and background distributions normalised 
to the results of the maximum-likelihood fit described in \Section{sec:Yields}. The cuts 
that define the signal region are indicated for each of the variables. The multijet background
estimate shown in the figure is discussed in \Section{sec:Backgrounds}.

%% file: Backgrounds.tex
\section{Background normalisation and modelling}
\label{sec:Backgrounds}

The largest background contributions to $t$-channel single top-quark production arise from 
\ttbar\ and $W$+jets production. The former is difficult to distinguish from the signal
since \ttbar\ events contain real top quarks in the final state. The $W$+jets production
contributes to the background if there is a $b$-quark in the final state or due to 
mistagging of jets containing other quark flavours. Multijet production via the strong 
interaction can contribute as well if, in addition to two reconstructed jets, an extra 
jet is misidentified as an isolated lepton, or if a non-prompt lepton appears to be 
isolated (both referred to as fake leptons). Other minor backgrounds originate from $Wt$, 
$s$-channel single top-quark, $Z$+jets and diboson production.

For all processes, except multijet production, the normalisation is initially estimated by 
using the Monte Carlo simulation scaled to the theoretical cross-section predictions, and the 
event distribution modelling is taken from simulation.

The \ttbar\ production cross-section is calculated at NNLO in QCD including resummation 
of next-to-next-to-leading-logarithm (NNLL) soft gluon terms with 
Top{\scriptsize ++}2.0~\cite{Cacciari:2011,Baernreuther:2012,Czakon:2012a,Czakon:2012b,Czakon:2013,Czakon:2011}. 
Its predicted value is $253^{+13}_{-15}$~pb~\cite{Cacciari:2011}. The quoted uncertainties include
the PDF and $\alpha_\mathrm{s}$ uncertainties calculated according to the PDF4LHC 
prescription~\cite{PDF4LHC} with the MSTW2008 NNLO~\cite{PDF_Martin,PDF_Martin_2}, CT10 NNLO~\cite{PDF_Lai,PDF_Gao} 
and NNPDF2.3 5f FFN~\cite{PDF_Ball} PDF sets, and the QCD scale uncertainty. The $t$-channel, 
$Wt$ and $s$-channel single-top-quark production cross-sections are calculated 
at NLO precision in QCD through NNLL resummation, leading to $87.7^{+3.4}_{-1.9}$~pb~\cite{Kidonakis_tchan}, 
$22.4\pm1.5$~pb~\cite{Kidonakis_Wt} and $5.6\pm0.2$~pb~\cite{Kidonakis_schan}, respectively. 
The calculations assume a top-quark mass of 172.5~GeV and use the MSTW2008 NNLO~\cite{PDF_Martin} 
PDFs. The quoted uncertainties include those due to the QCD scale uncertainty and the correlated 
PDF--$\alpha_{\mathrm{s}}$ uncertainty. 

The cross-sections for inclusive $W$- and $Z$-boson production are estimated with NNLO precision
using the FEWZ program~\cite{FEWZ_1,FEWZ_2} and the MSTW2008 NNLO PDFs. The diboson samples are 
normalised to the NLO cross-section predictions calculated with MCFM~\cite{MCFM}. A normalisation 
uncertainty of 20\% is assigned to the $W$+jets background. This uncertainty is estimated from 
parameter variations of the \textsc{Sherpa} generator covering the measured $W$+jets cross-sections~\cite{ATLAS_Wjets}. 
A normalisation uncertainty of 20\% is also assumed for the $Z$+jets and diboson processes.

The normalisation as well as the event modelling of the multijet background is estimated from 
data using the matrix method~\cite{ATLAS_topreco,ATLAS_ttbar}. This method allows the derivation 
of the true composition of the data sample in terms of prompt (real) and fake leptons from its 
observed composition in terms of tight (signal selection) and loose leptons. An alternative 
normalisation and modelling based on the mixed data--simulation jet-electron method~\cite{ATLAS_tchan_letter,ATLAS_tchan} 
and the purely data-driven anti-muon selection~\cite{ATLAS_topreco} are used to estimate the 
systematic uncertainties. From the comparison an overall normalisation uncertainty of 70\% is 
assigned to the multijet contribution.

To check the modelling of the \ttbar\ and $W$+jets background contributions, the simulated 
events are compared to the data in two dedicated background-dominated regions. Samples enriched 
in \ttbar\ events (\ttbar\ control region) are defined by considering events preselected as
explained in \Section{sec:Selections}, but containing two additional jets that are required to 
be untagged. This control region is also used in the normalisation fit described in \Section{sec:Yields}. 
Samples enriched in $W$+jets events ($W$+jets control region) are selected by applying a 
relaxed $b$-tagging requirement corresponding to an efficiency of 80\%. In addition, all events
satisfying the signal $b$-tagging requirement are excluded. For these two control regions 
the dilepton rejection and the four final selection cuts are not applied. An additional category 
of events is defined by selecting all events not passing the four signal selection cuts 
(anti-signal region). This region is only used in the normalisation fit, in combination with 
the \ttbar\ control region. It is preferred to the $W$+jets control region to constrain the
$W$+jets normalisation because it has a flavour composition more similar to that of the signal 
region. The predicted fraction of heavy-flavour events in the $W$+jets contribution is around 
95\% for both the signal and anti-signal selections, whereas it is 55\% in the $W$+jets control 
region.

\begin{figure}[t]
\captionsetup[subfloat]{farskip=2pt,captionskip=1pt}
\centering
\subfloat[]
{\includegraphics[width=0.48\textwidth]{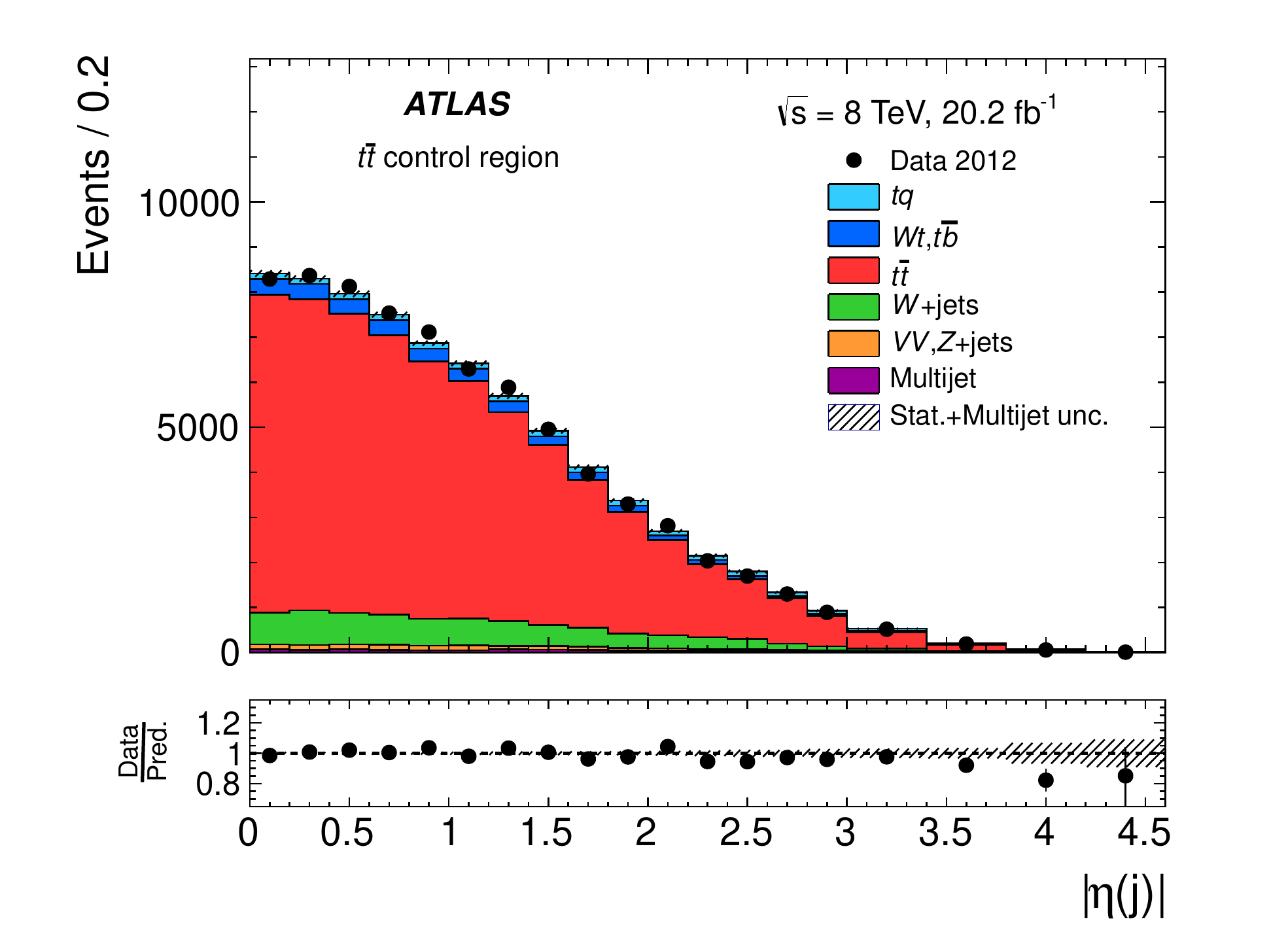}
\label{fig:ttbar_abs_ujet_eta}}
\subfloat[]
{\includegraphics[width=0.48\textwidth]{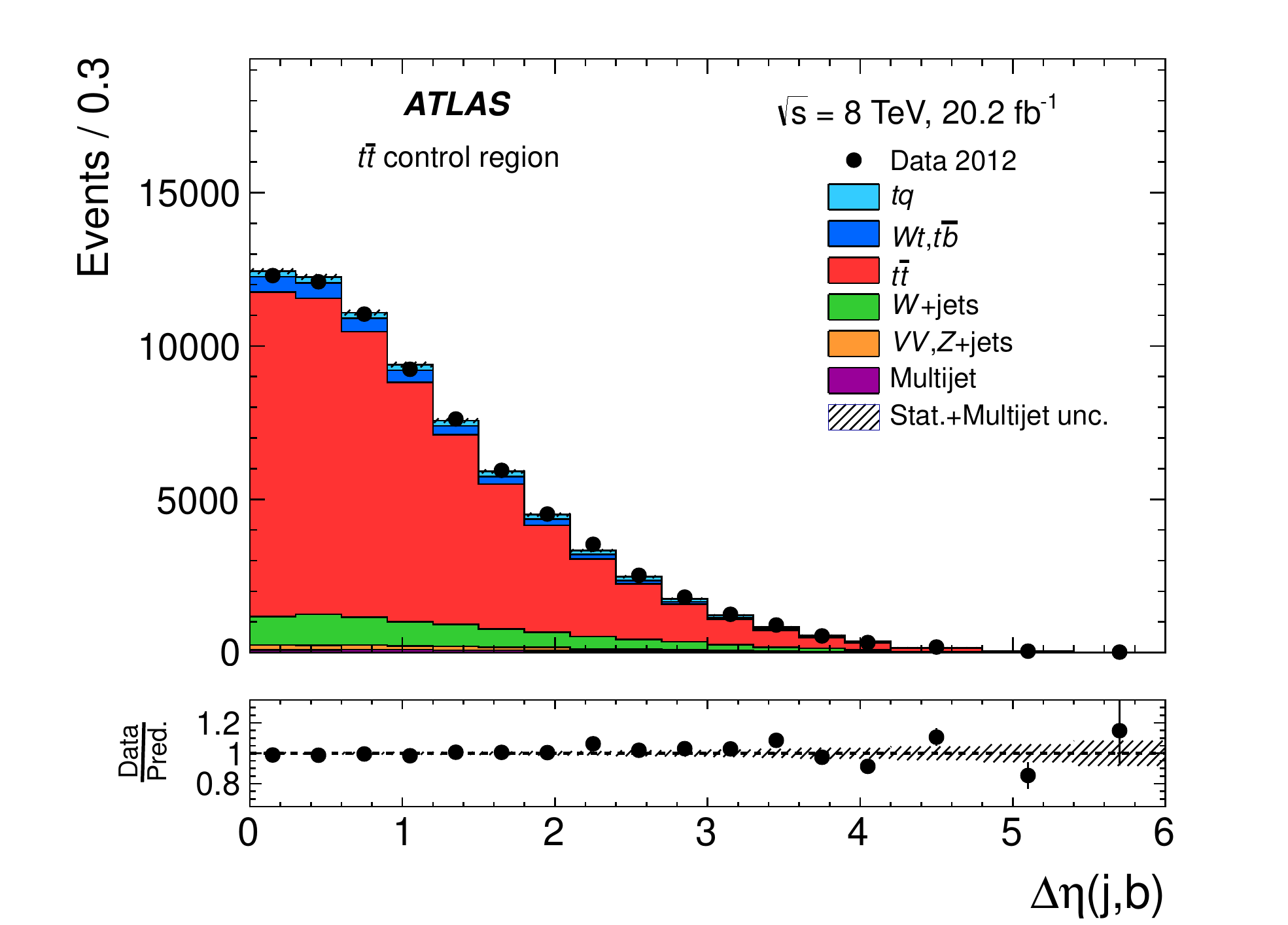}
\label{fig:ttbar_dEta_bjet_ujet}}
\hfill
\subfloat[]
{\includegraphics[width=0.48\textwidth]{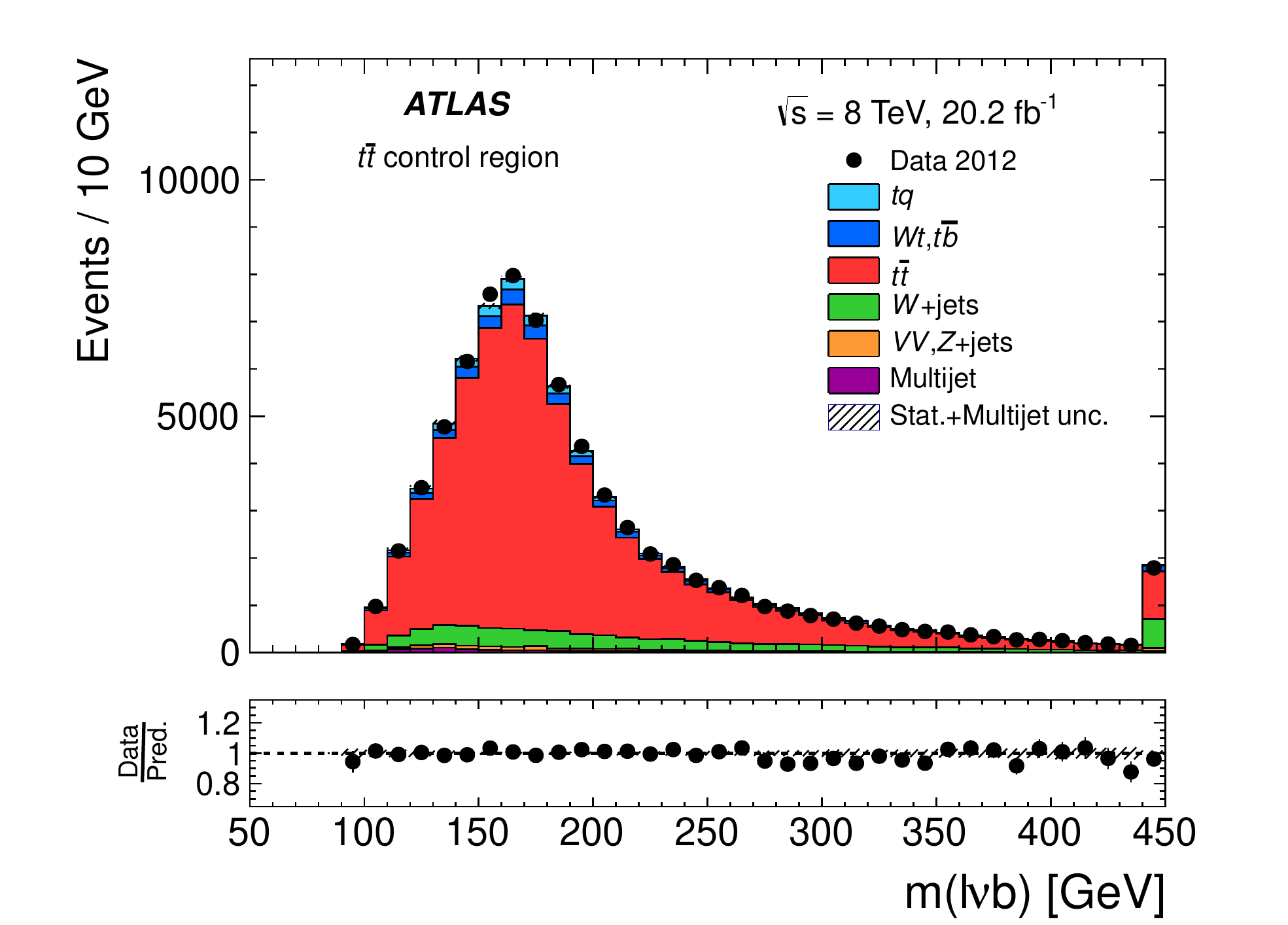}
\label{fig:ttbar_top_m}}
\subfloat[]
{\includegraphics[width=0.48\textwidth]{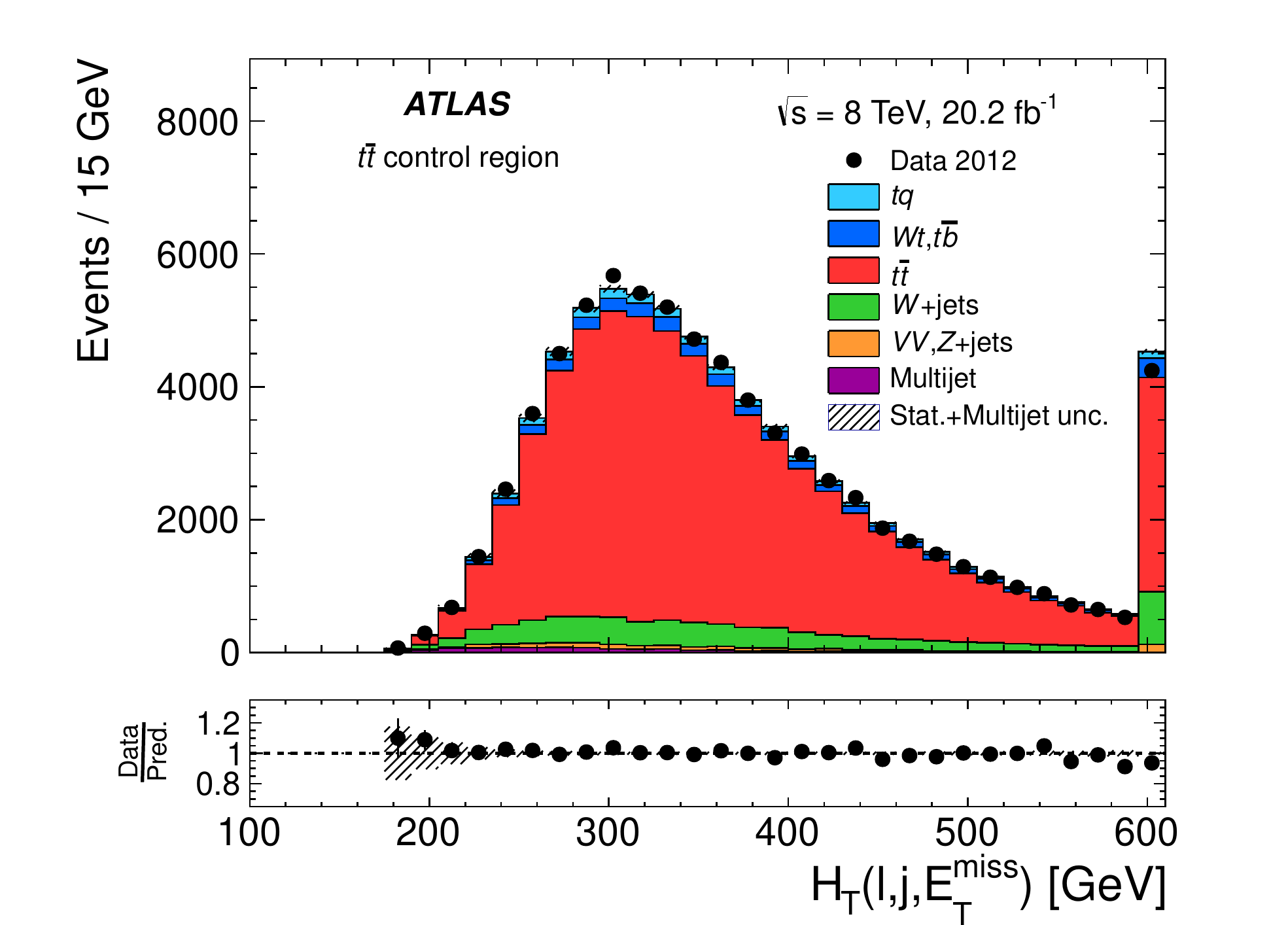}
\label{fig:ttbar_Ht_event}}
\caption
{Distributions of the selection variables in the \ttbar\ control region:
\protect\subref{fig:ttbar_abs_ujet_eta} $|\eta|$ of the untagged jet,
\protect\subref{fig:ttbar_dEta_bjet_ujet} separation in $\eta$ between the untagged and $b$-tagged jets,
\protect\subref{fig:ttbar_top_m} reconstructed top-quark mass, and
\protect\subref{fig:ttbar_Ht_event} scalar sum of the \pT\ of the lepton, the \pT\ of the jets and \MET.
The observed distributions are compared to the predicted signal and background distributions, normalised 
to the results of the maximum-likelihood fit. The labels $tq$ and $t\bar{b}$ refer to the $t$-channel 
and $s$-channel single-top-quark processes, respectively, and $VV$ to diboson production. The uncertainty 
bands include the statistical post-fit uncertainty, the uncertainty due to the limited size of the simulation 
samples and the uncertainty in the normalisation of the multijet background, added in quadrature. The last bin 
of the histograms includes overflows. The lower panels show the ratio of data to prediction.}
\label{fig:ttbar_selections}
\end{figure}

\begin{figure}[t]
\captionsetup[subfloat]{farskip=2pt,captionskip=1pt}
\centering
\subfloat[]
{\includegraphics[width=0.48\textwidth]{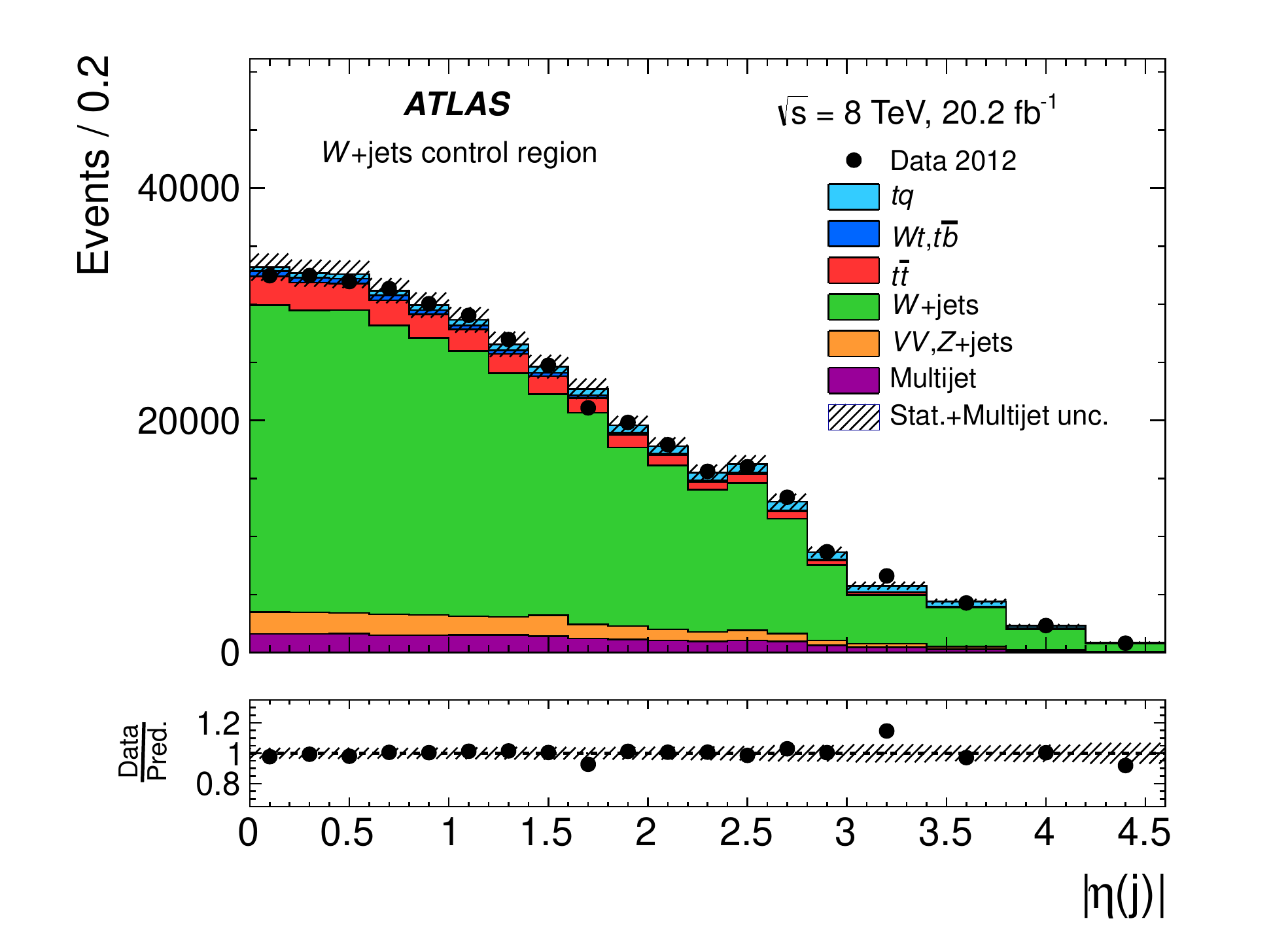}
\label{fig:Wjets_abs_ujet_eta}}
\subfloat[]
{\includegraphics[width=0.48\textwidth]{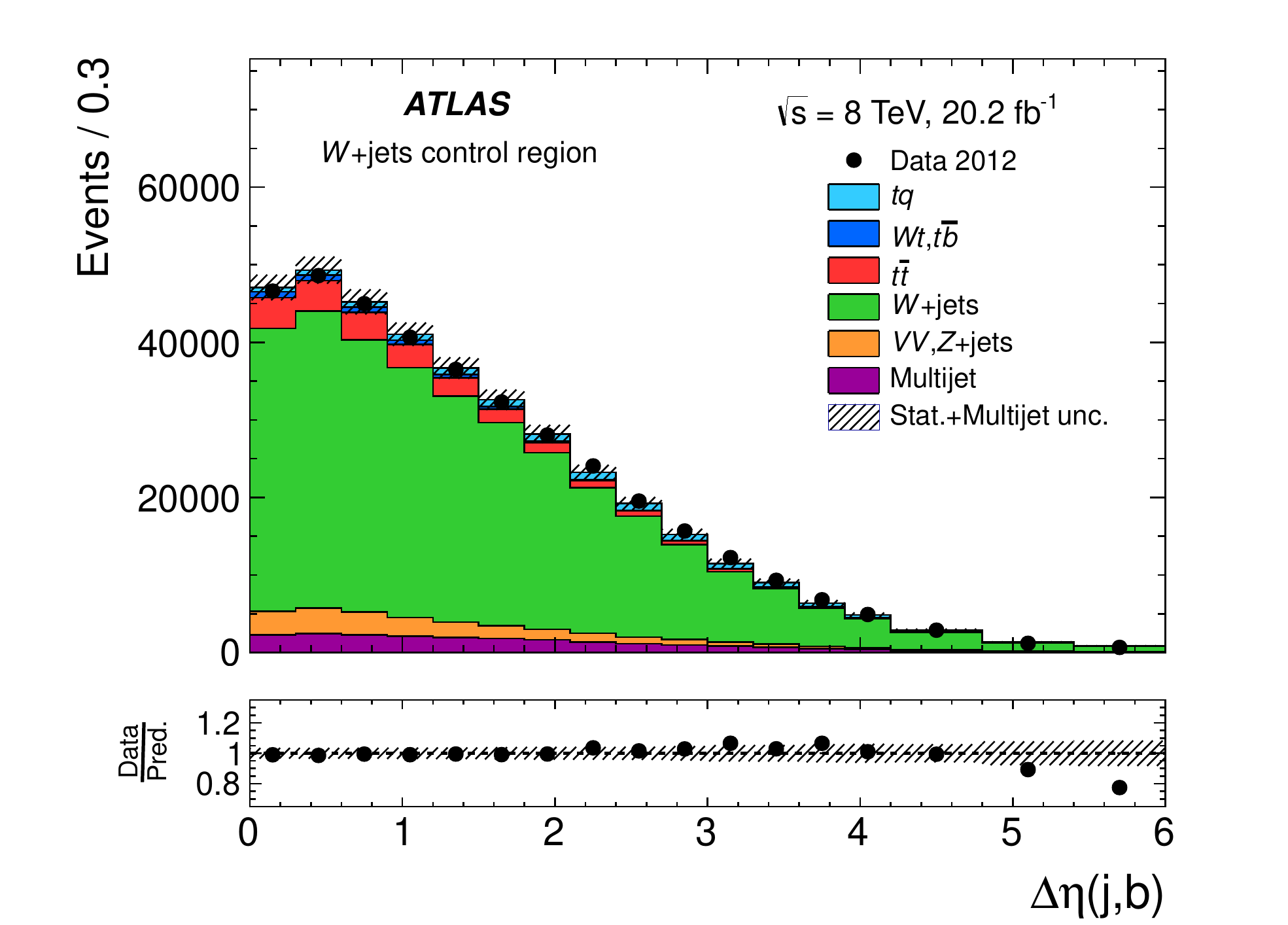}
\label{fig:Wjets_dEta_bjet_ujet}}
\hfill
\subfloat[]
{\includegraphics[width=0.48\textwidth]{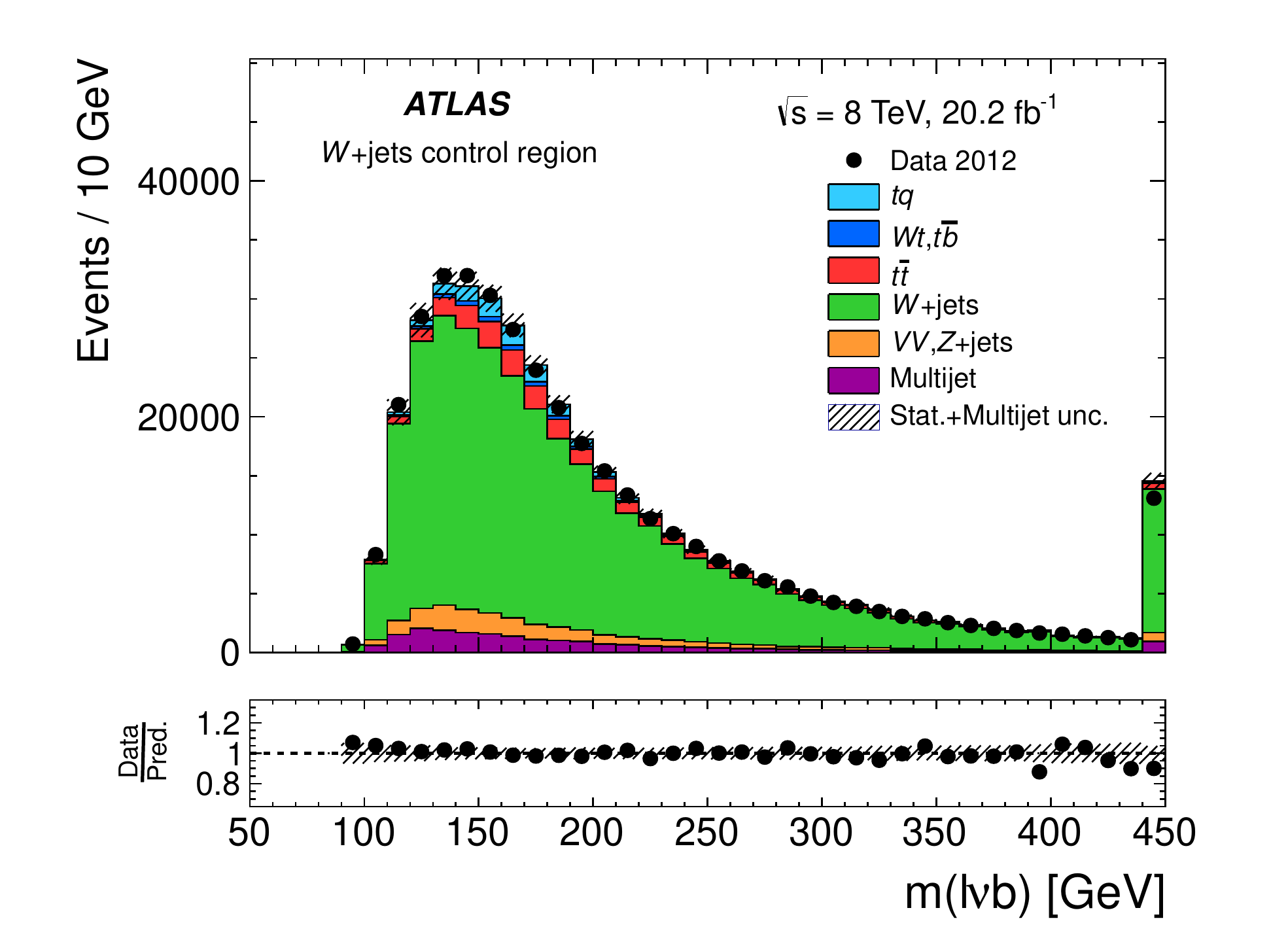}
\label{fig:Wjets_top_m}}
\subfloat[]
{\includegraphics[width=0.48\textwidth]{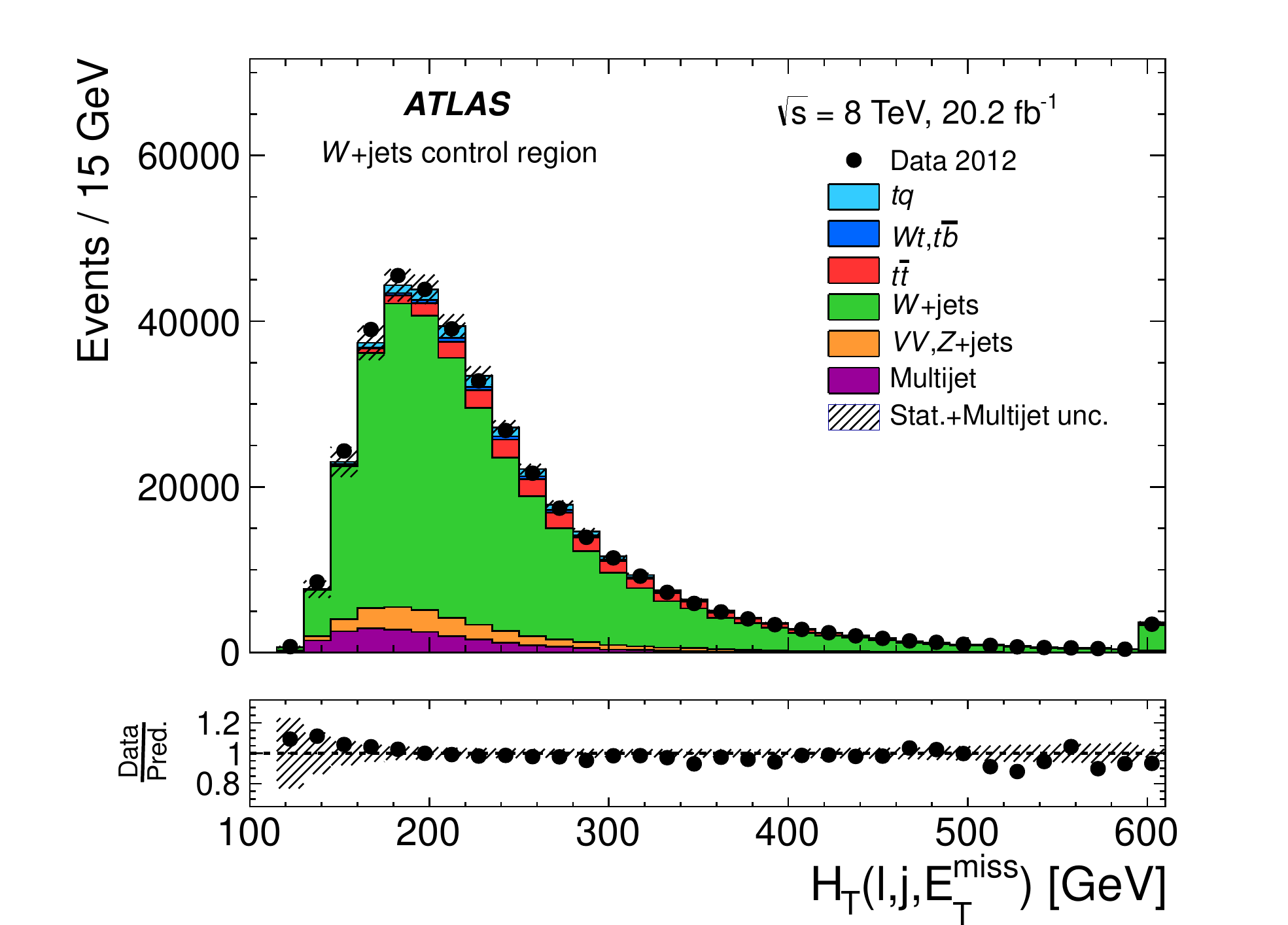}
\label{fig:Wjets_Ht_event}}
\caption
{Distributions of the selection variables in the $W$+jets control region:
\protect\subref{fig:Wjets_abs_ujet_eta} $|\eta|$ of the untagged jet,
\protect\subref{fig:Wjets_dEta_bjet_ujet} separation in $\eta$ between the untagged and $b$-tagged jets,
\protect\subref{fig:Wjets_top_m} reconstructed top-quark mass, and
\protect\subref{fig:Wjets_Ht_event} scalar sum of the \pT\ of the lepton, the \pT\ of the jets and \MET.
The observed distributions are compared to the predicted signal and background distributions. 
The $W$+jets distributions are normalised to match the observed number of events. The labels 
$tq$ and $t\bar{b}$ refer to the $t$-channel and $s$-channel single-top-quark processes, respectively, 
and $VV$ to diboson production. The uncertainty bands include the uncertainty due to the limited size 
of the simulation samples and the uncertainty in the normalisation of the multijet background, added in
quadrature. The last bin of the histograms includes overflows. The lower panels show the ratio of data 
to prediction.}
\label{fig:Wjets_selections}
\end{figure}

Good overall data--prediction agreement is found in the \ttbar\, $W$+jets and anti-signal control 
regions for the relevant kinematic observables, as well as for the various angular observables used 
in the measurements. Figure~\ref{fig:ttbar_selections} shows the distributions in the \ttbar\ 
control region of the four variables used to define the final selections. The distributions 
obtained in the $W$+jets control region are displayed in Figure~\ref{fig:Wjets_selections}.

%% file: Yields.tex
\section{Signal and background event yields}
\label{sec:Yields}

The signal and background event yields are estimated through a simultaneous
maximum-likelihood fit to the numbers of data events observed in the signal 
and anti-signal regions, and in the \ttbar\ control region.

The likelihood function~\cite{ATLAS_tchan} is given by the product of Poisson 
probability terms associated with the fitted regions, combined with the product of 
Gaussian priors to constrain the background rates to their predictions within the 
associated uncertainties. In the fit the $t$-channel single-top-quark contribution 
is treated as unconstrained. The top-quark background contributions (\ttbar, $Wt$ 
and $s$-channel single top-quark production) are merged with their relative fractions 
taken from simulation, and the applied constraint is derived from the combination 
of their cross-section uncertainties presented in \Section{sec:Backgrounds}. The 
flavour composition of the $W$+jets contribution is taken from
simulation. In all fitted regions the production of a $W$ boson in association with
heavy-flavour jets is the dominant contribution to the $W$+jets background, predicted 
to be around 95\% in the three regions. The $Z$+jets and diboson contributions, which 
are very low in the signal region (2\% of the total expectation), are merged and 
fixed to the predictions. The multijet contribution is kept fixed to its data-driven 
estimate.

The results of the maximum-likelihood fit together with the associated statistical
uncertainties (referred to as statistical post-fit uncertainties) are shown in 
Table~\ref{tab:Scales}. They are presented as scale factors to be applied to the 
predicted event yields. The results are found to be stable when the constraints 
imposed on the top-quark and $W$+jets backgrounds are significantly relaxed. Table~\ref{tab:Yields} 
provides the signal and background event yields in the signal region after scaling 
to the results of the fit to the data. The signal-to-background ratio is 1.2, the 
$t$-channel single top-quark production representing 54\% of the total expectation. 
The two main background contributions come from $W$+jets (19\%) and \ttbar\ production 
(18\%).

\begin{table}[!ht]
\begin{center}
\begin{tabular}{lr@{$\,\pm\,$}l}
\toprule
Process                       & \multicolumn{2}{c}{Scale factor} \\
\midrule
$t$-channel                   & 0.95 & 0.02 \\
$t\bar{t}$, $Wt$, $s$-channel & 1.01 & 0.01 \\
$W$+jets                      & 1.10 & 0.01 \\
\bottomrule
\end{tabular}
\caption{Scale factors and uncertainties extracted for the signal and background processes 
from the simultaneous maximum-likelihood fit of the event yields in the signal, anti-signal
and \ttbar\ regions. The quoted uncertainties are statistical only.}
\label{tab:Scales}
\end{center}
\end{table} 

\begin{table}[!ht]
\begin{center}
\begin{tabular}{lr@{$\,\pm\,$}l}
\toprule
Process                                 & \multicolumn{2}{c}{Event yield} \\
\midrule
$t$-channel                             & 5700  & 110  \\
$Wt$, $s$-channel                       & 265   & 12   \\
$t\bar{t}$                              & 1914  & 15   \\
$W$+jets                                & 2044  & 57   \\
$Z$+jets, diboson                       & 188   & 9    \\
Multijet                                & 420   & 290  \\
\midrule
Total expectation                       & 10530 & 320  \\
Data                                    & \multicolumn{2}{l}{10527} \\
\bottomrule
\end{tabular}
\caption{Signal and background event yields in the signal region after scaling to the 
results of the maximum-likelihood fit. The quoted uncertainties add in quadrature the 
post-fit uncertainties and the uncertainties due to the limited size of the simulation 
samples, except for the data-driven multijet contribution to which the normalisation 
uncertainty of 70\% is applied. The total expectation is compared to the observed number 
of events.}
\label{tab:Yields}
\end{center}
\end{table}

%% file: Distributions.tex
\section{Angular distributions}
\label{sec:Distributions}

\begin{figure}[t]
\captionsetup[subfloat]{farskip=2pt,captionskip=1pt}
\centering
\subfloat[]
{\includegraphics[width=0.48\textwidth]{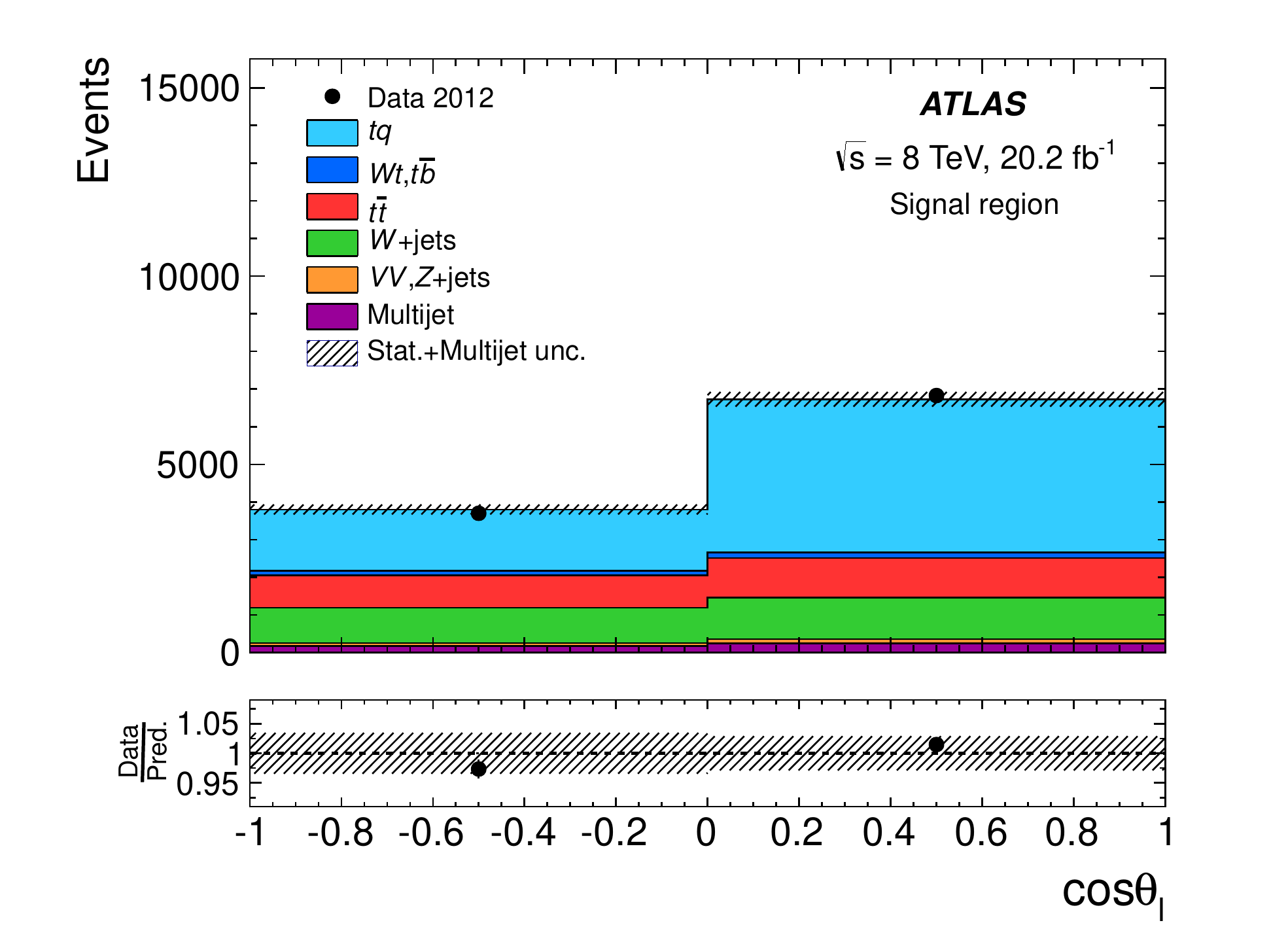}
\label{fig:Polar_cos_lepton_ujet_tchan_1tag_cuts_noFiducial_2jets_Leptons_2}}
\subfloat[]
{\includegraphics[width=0.48\textwidth]{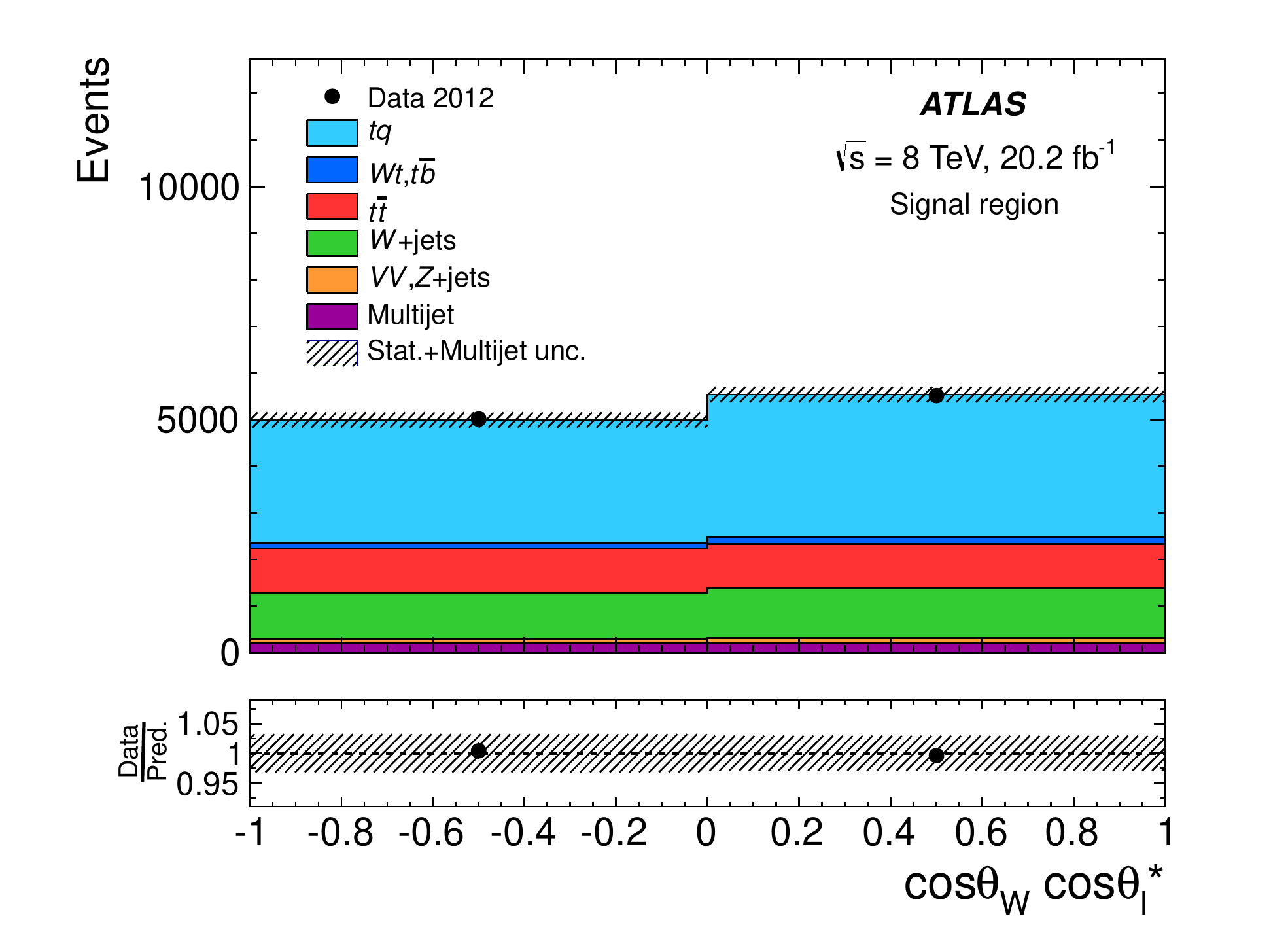}
\label{fig:Polar_cos_lepton_tW_tchan_1tag_cuts_noFiducial_2jets_Leptons_2}}
\hfill
\subfloat[]
{\includegraphics[width=0.48\textwidth]{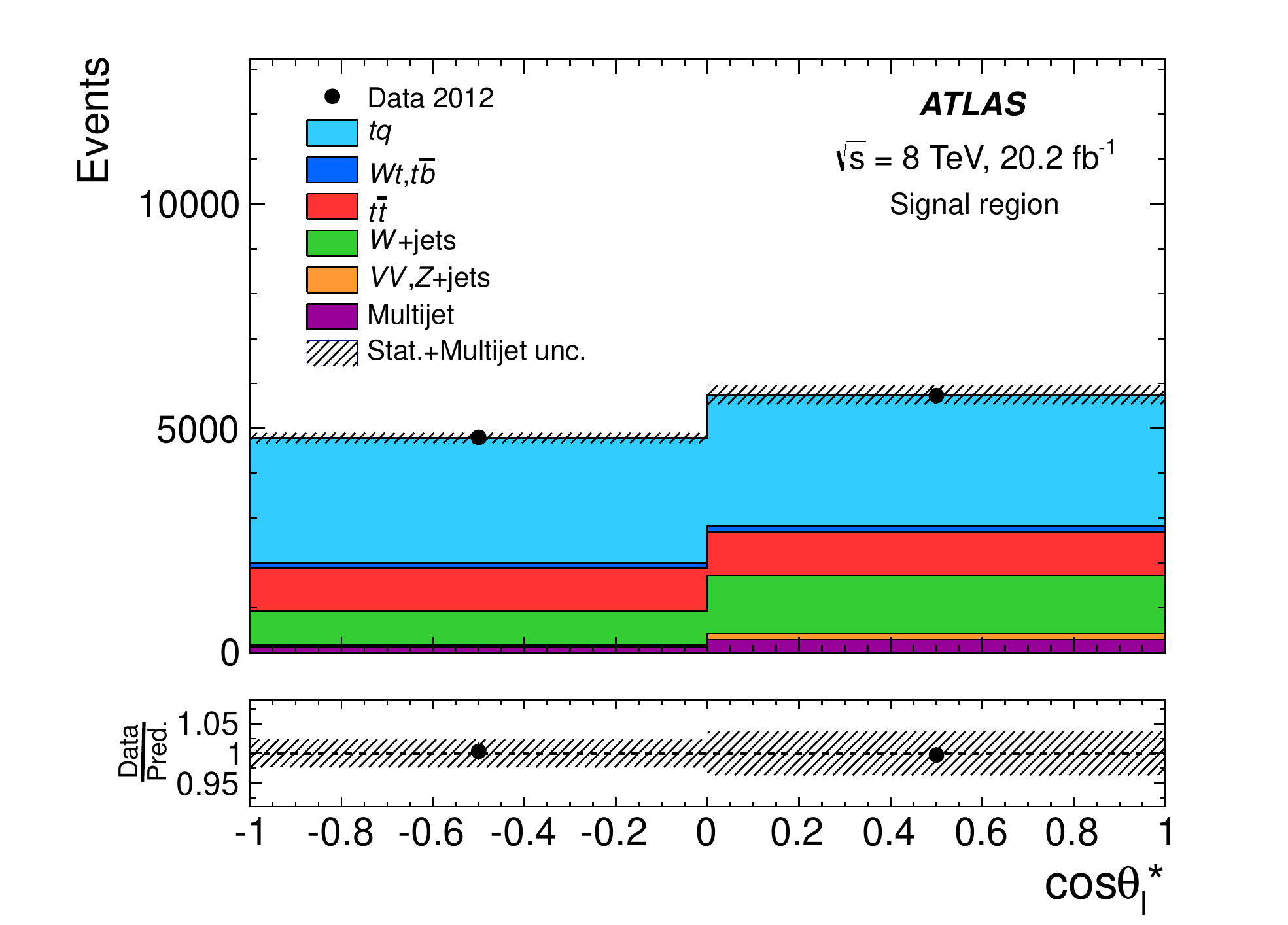}
\label{fig:Polar_cos_lepton_bjet_tchan_1tag_cuts_noFiducial_2jets_Leptons_2}}
\subfloat[]
{\includegraphics[width=0.48\textwidth]{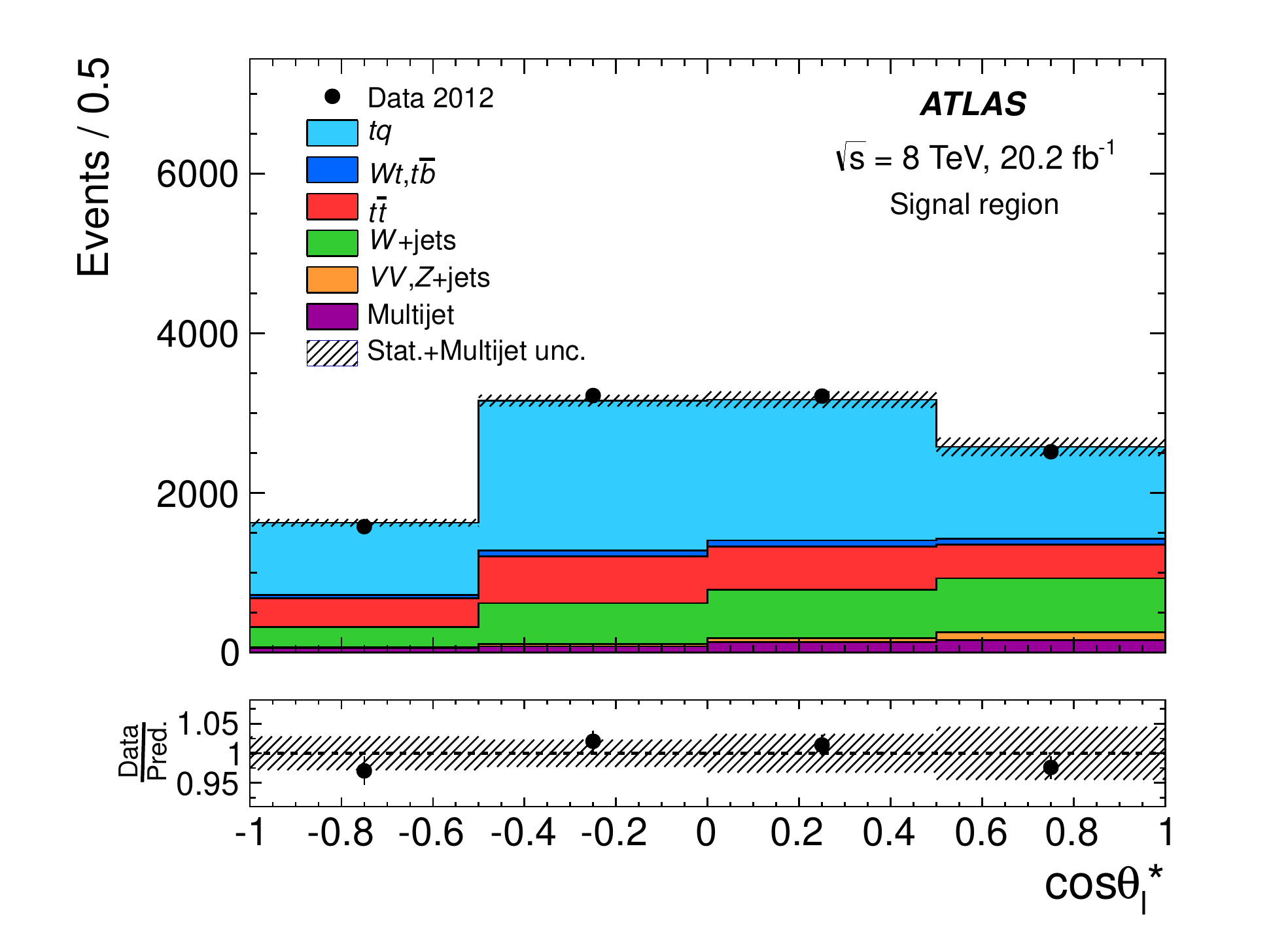}
\label{fig:Polar_cos_lepton_bjet_tchan_1tag_cuts_noFiducial_2jets_Leptons_4}}
\caption
{Distributions in the signal region of the angular observables used to measure the various asymmetries:
\protect\subref{fig:Polar_cos_lepton_ujet_tchan_1tag_cuts_noFiducial_2jets_Leptons_2} $\cos\theta_{\ell}$ for $\AFB^{\ell}$,
\protect\subref{fig:Polar_cos_lepton_tW_tchan_1tag_cuts_noFiducial_2jets_Leptons_2} $\cos\theta_W \cos\theta_{\ell}^*$ for $\AFB^{tW}$,
\protect\subref{fig:Polar_cos_lepton_bjet_tchan_1tag_cuts_noFiducial_2jets_Leptons_2} $\cos\theta_{\ell}^*$ with two bins for $\AFB$, and
\protect\subref{fig:Polar_cos_lepton_bjet_tchan_1tag_cuts_noFiducial_2jets_Leptons_4} $\cos\theta_{\ell}^*$ with four bins for $\AEC$.
The observed distributions are compared to the predicted signal and background distributions,
normalised to the results of the maximum-likelihood fit. The template $t$-channel distributions 
are taken from the baseline \textsc{Powheg-Box} sample. The labels $tq$ and $t\bar{b}$ refer to 
the $t$-channel and $s$-channel single-top-quark processes, respectively, and $VV$ to diboson 
production. The uncertainty bands include the statistical post-fit uncertainty, the uncertainty 
due to the limited size of the simulation samples and the uncertainty in the normalisation of the 
multijet background, added in quadrature. The lower panels show the ratio of data to prediction.}
\label{fig:Distributions_1}
\end{figure}

\begin{figure}[t]
\captionsetup[subfloat]{farskip=2pt,captionskip=1pt}
\centering
\subfloat[]
{\includegraphics[width=0.48\textwidth]{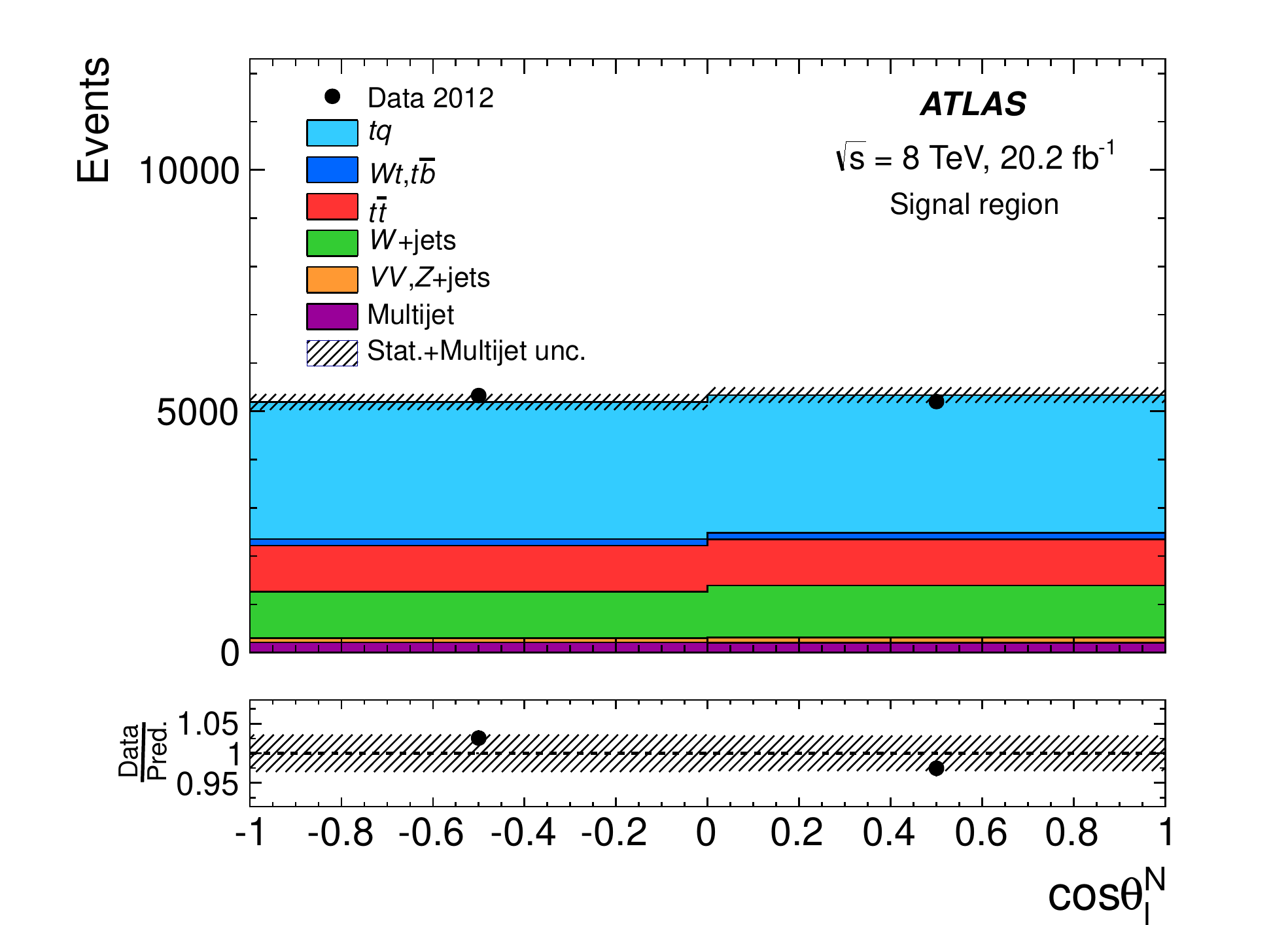}
\label{fig:Polar_cos_lepton_normal_tchan_1tag_cuts_noFiducial_2jets_Leptons_2}}
\subfloat[]
{\includegraphics[width=0.48\textwidth]{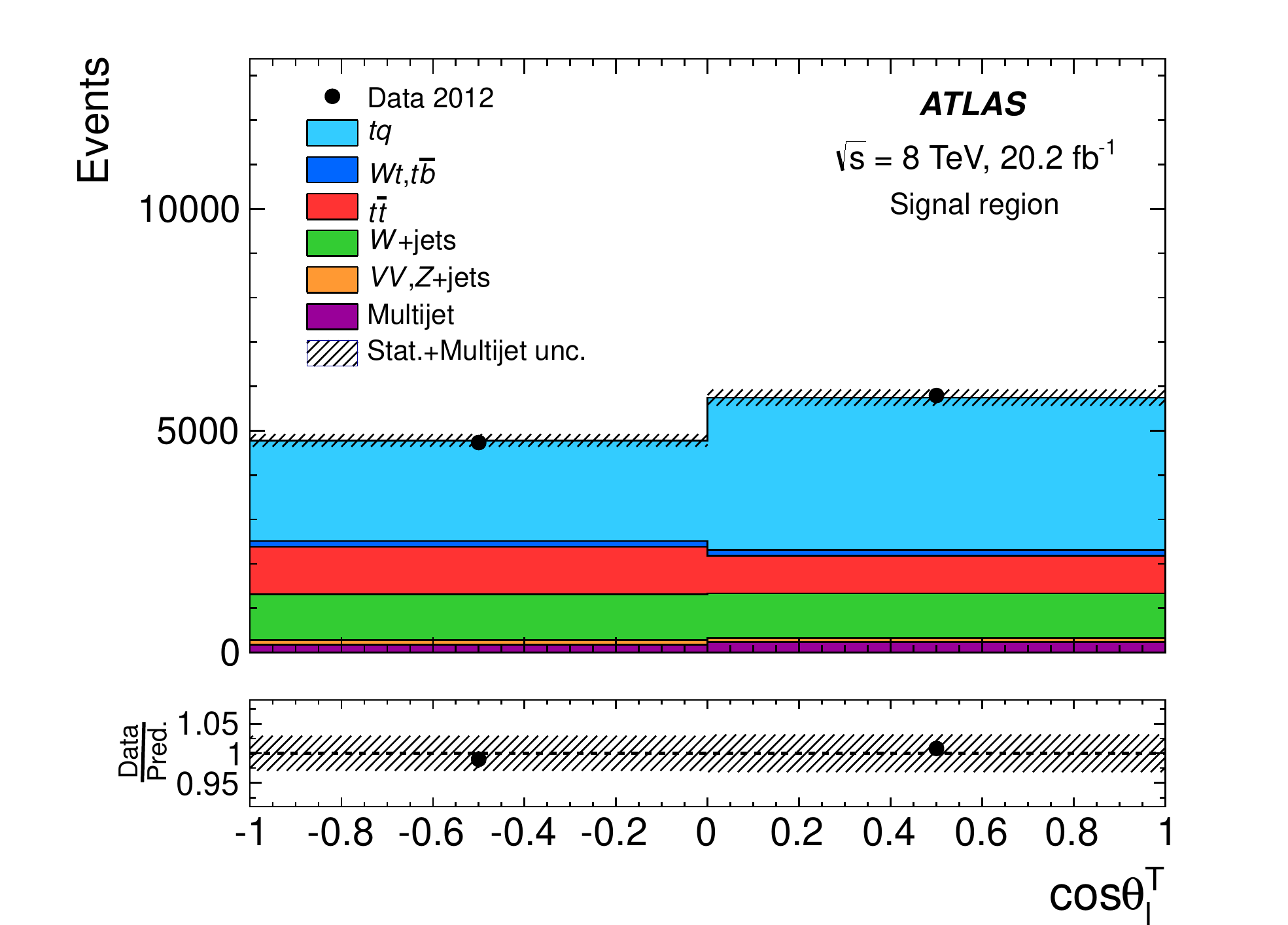}
\label{fig:Polar_cos_lepton_trans_tchan_1tag_cuts_noFiducial_2jets_Leptons_2}}
\hfill
\subfloat[]
{\includegraphics[width=0.48\textwidth]{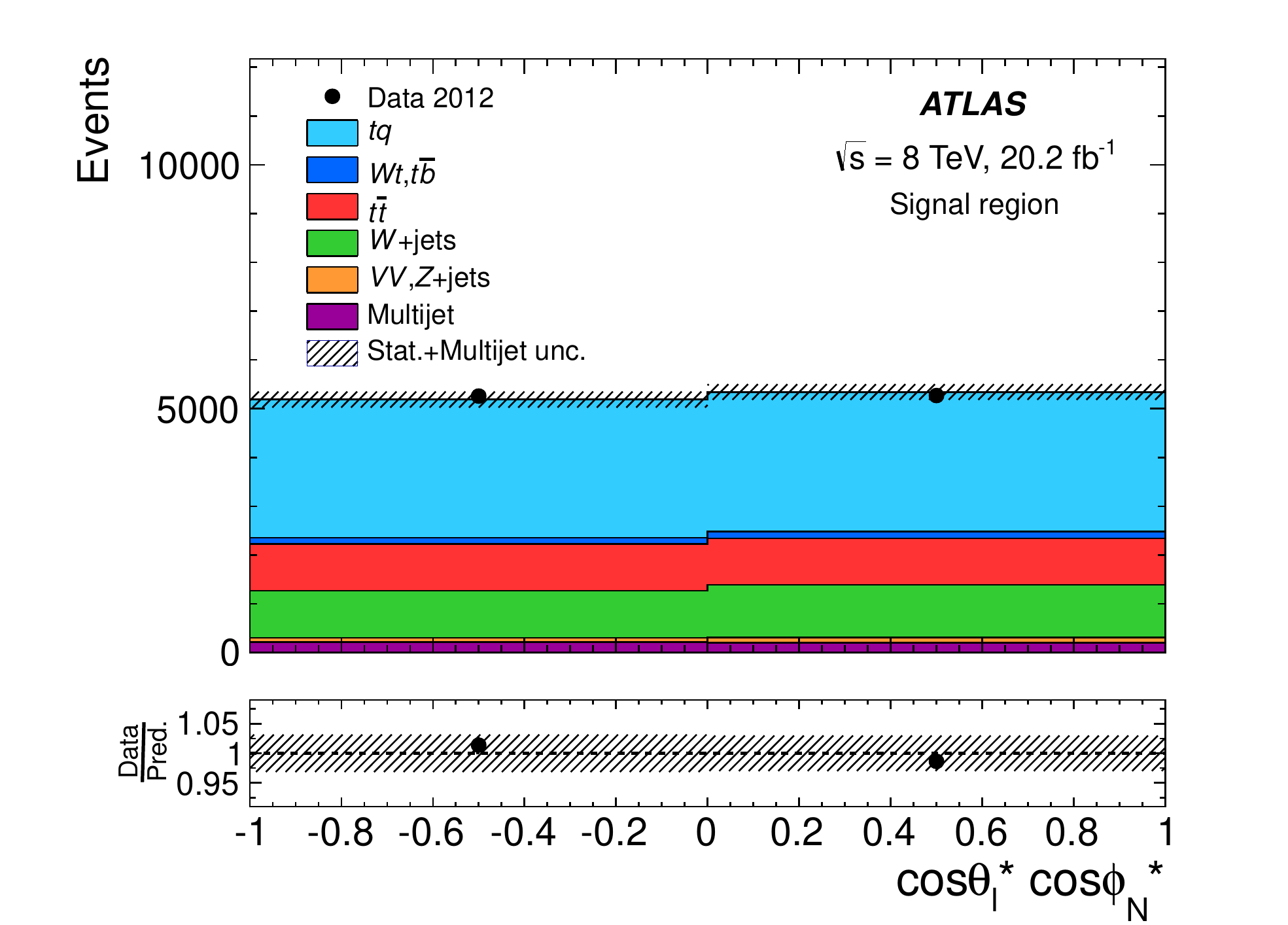}
\label{fig:Polar_cos_lepton_bjet_n_tchan_1tag_cuts_noFiducial_2jets_Leptons_2}}
\subfloat[]
{\includegraphics[width=0.48\textwidth]{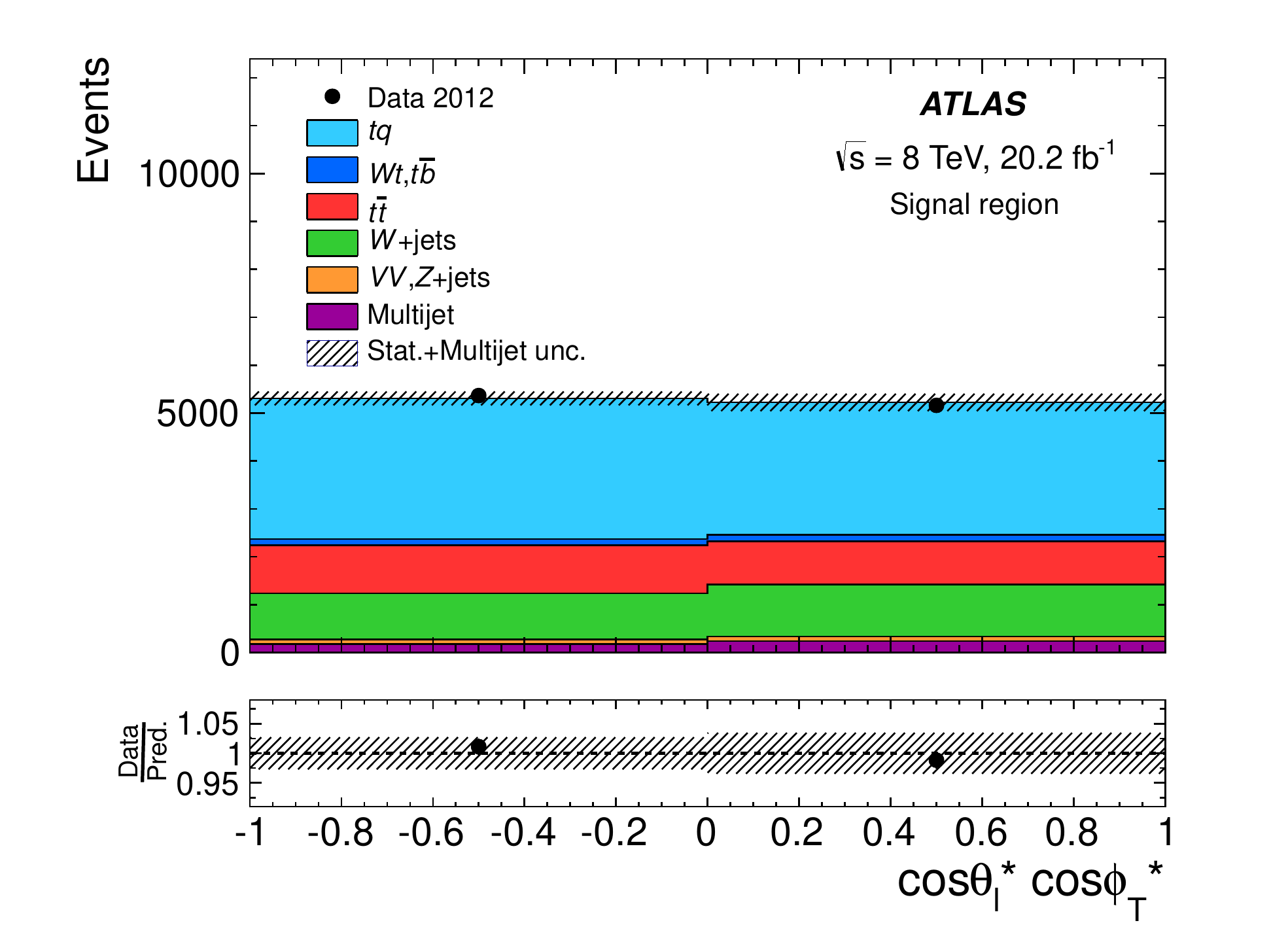}
\label{fig:Polar_cos_lepton_bjet_t_tchan_1tag_cuts_noFiducial_2jets_Leptons_2}}
\caption
{Distributions in the signal region of the angular observables used to measure the various asymmetries:
\protect\subref{fig:Polar_cos_lepton_normal_tchan_1tag_cuts_noFiducial_2jets_Leptons_2} $\cos\theta_{\ell}^{N}$ for $\AFB^{N}$,
\protect\subref{fig:Polar_cos_lepton_trans_tchan_1tag_cuts_noFiducial_2jets_Leptons_2} $\cos\theta_{\ell}^{T}$ for $\AFB^{T}$,
\protect\subref{fig:Polar_cos_lepton_bjet_n_tchan_1tag_cuts_noFiducial_2jets_Leptons_2} $\cos\theta_{\ell}^* \cos\phi_{N}^*$ for $\AFB^{N,\phi}$, and
\protect\subref{fig:Polar_cos_lepton_bjet_t_tchan_1tag_cuts_noFiducial_2jets_Leptons_2} $\cos\theta_{\ell}^* \cos\phi_{T}^*$ for $\AFB^{T,\phi}$.
The observed distributions are compared to the predicted signal and background distributions,
normalised to the results of the maximum-likelihood fit. The template $t$-channel distributions 
are taken from the baseline \textsc{Powheg-Box} sample. The labels $tq$ and $t\bar{b}$ refer to 
the $t$-channel and $s$-channel single-top-quark processes, respectively, and $VV$ to diboson 
production. The uncertainty bands include the statistical post-fit uncertainty, the uncertainty due 
to the limited size of the simulation samples and the uncertainty in the normalisation of the multijet 
background, added in quadrature. The lower panels show the ratio of data to prediction.}
\label{fig:Distributions_2}
\end{figure}

The distributions observed at reconstruction level for the angular observables used to measure 
the various asymmetries are shown in Figures~\ref{fig:Distributions_1} and \ref{fig:Distributions_2}. 
They are compared to the predicted signal and background distributions, normalised to the results of 
the maximum-likelihood fit. To minimise the unfolding corrections that are applied after background
subtraction, two bins are chosen for the angular distributions from which forward-backward asymmetries 
are extracted, while four bins are used for the angular distribution from which the $\AEC$ asymmetry 
is determined.

Depending on the angular observable, as described in Section~\ref{sec:Observables}, the charged-lepton 
four-momentum is computed in the rest frame of the reconstructed top quark or in the rest frame
of the reconstructed $W$ boson. The angular observables related to the top-quark polarisation are 
defined by taking the momentum of the untagged jet as the spectator-quark direction, whereas those 
related to the $W$-boson spin observables are defined by considering the reverse momentum of the 
$b$-tagged jet as the $W$-boson direction.

%% file: Unfolding.tex
\section{Unfolding}
\label{sec:Unfolding}

The measured angular distributions are unfolded to the parton level,\footnote{Partons 
are defined from the matrix-element hard process and immediate decays.} so that 
the asymmetries extracted from the corrected angular distributions can be directly 
compared to theoretical calculations. The unfolding corrections account for 
distortions due to detector resolution, selection efficiencies, and reconstruction 
of the $W$ boson and top quark. They also include the effects due to hadronisation 
and parton showering. 

The unfolding procedure is applied to the angular distributions after subtracting the 
background contributions, and is based on a matrix inversion combined with an efficiency 
correction. The number of unfolded signal events $N^{\text{unfolded}}_{j}$ in each bin $j$ 
of the parton-level distribution is obtained from the background-subtracted yields 
$N^{\text{measured}}_{i}$ measured in all bins $i$ of the reconstructed distribution, 
according to

\begin{equation}
N^{\text{unfolded}}_{j} = \frac{\sum_{i} M^{-1}_{ji}N^{\text{measured}}_{i}}{\epsilon_{j}} ,\\
\label{eq:asymmetry}
\end{equation}
 
\noindent
where $M_{ji}$ is the migration matrix which relates the parton-level and reconstructed 
values of the considered angular variable, and $\epsilon_{j}$ is the event selection 
efficiency. Both the migration matrix and the selection efficiency are computed using samples 
of $t$-channel events simulated with the \textsc{Protos} generator, as described below. 
For the chosen numbers of bins, the fractions of simulated events belonging to the diagonal 
elements of the migration matrices are found to be between 68\% and 90\%, depending on the angular
observable. The selection efficiencies are between 0.6\% and 1.6\%, depending on the angular 
observable and on the bin range. The matrix inversion is performed by using the iterative 
Bayesian method~\cite{BayesMethod} as implemented in the RooUnfold framework~\cite{RooUnfold}. 
The number of iterations is chosen such that the absolute change in the extracted asymmetry 
between two successive steps becomes lower than 0.0005. The unfolding procedure has been 
validated through convergence and closure tests performed by using template distributions 
constructed from the $t$-channel \textsc{Powheg-Box} and \textsc{Protos} samples presented 
in Section~\ref{sec:Samples}. The closure tests showed that the residual bias induced by 
the unfolding method is negligible, whatever the measured asymmetry.

With the aim of testing their compatibility with the Standard Model predictions, all asymmetries 
described in \Section{sec:Observables}, except $\AFB^{N}$, are extracted using the \textsc{Protos} 
simulation generated with the Standard Model values of the $Wtb$ couplings to determine the 
migration matrix and the selection efficiency. For all the asymmetry measurements, the Standard 
Model $Wtb$ couplings, as implemented in the \textsc{Powheg-Box} generator, are considered for the 
subtracted top-quark backgrounds.

To constrain Im\,\gr\ using the method explained in \Section{sec:Observables}, the $\AFB^{N}$
and $\AFB^{\ell}$ asymmetries must be measured without any assumption about Im\,\gr. 
It is observed that the presence of anomalous couplings in general modifies the kinematics 
in such a way that the efficiency corrections are dependent on the $Wtb$ couplings. While 
the measurement of $\AFB^{\ell}$ is found to be independent of the value of Im\,\gr\ assumed 
in the unfolding corrections, the measurement of $\AFB^{N}$ is found to depend on the unfolding 
corrections used. By applying an interpolation technique it is possible to unfold the 
$\cos\theta_{\ell}^N$ angular distribution independently of any assumption about Im\,\gr, 
so that the extracted $\AFB^{N}$ asymmetry, combined with $\AFB^{\ell}$, can be used 
to constrain this coupling.

The interpolation method is based on determining the unfolding corrections using a linear 
combination of the migration and efficiency corrections provided by five {\textsc Protos} 
samples in which Im\,\gr\ is varied (Im\,\gr\ $=0, \pm0.094, \pm0.23$). An iterative 
procedure is applied to determine the coefficients of the linear combination until convergence 
is reached in the extracted $\AFB^{N}$ asymmetry. The method proceeds as follows. An initial 
value of $\AFB^{N}$ is first extracted using the standard {\textsc Protos} unfolding corrections. 
This value is then used to determine, via a Lagrange interpolation, the weights to be 
applied to the five predicted corrections. A new value of $\AFB^{N}$ is obtained after 
unfolding the $\cos\theta_{\ell}^N$ angular distribution with these corrections using 
the Bayesian method. The chosen convergence criterion for the interpolation procedure 
requires that the difference between the extracted $\AFB^{N}$ from two successive steps 
is smaller than 0.0005. By using template distributions given by \textsc{Protos} samples 
not used in the linear combination of the unfolding corrections (Im\,\gr\ $=\pm0.043, \pm0.144$), 
it has been checked that this method recovers the generated asymmetries at parton level.

\begin{figure}[t]
\captionsetup[subfloat]{farskip=2pt,captionskip=1pt}
\centering
\subfloat[]
{\includegraphics[width=0.48\textwidth]{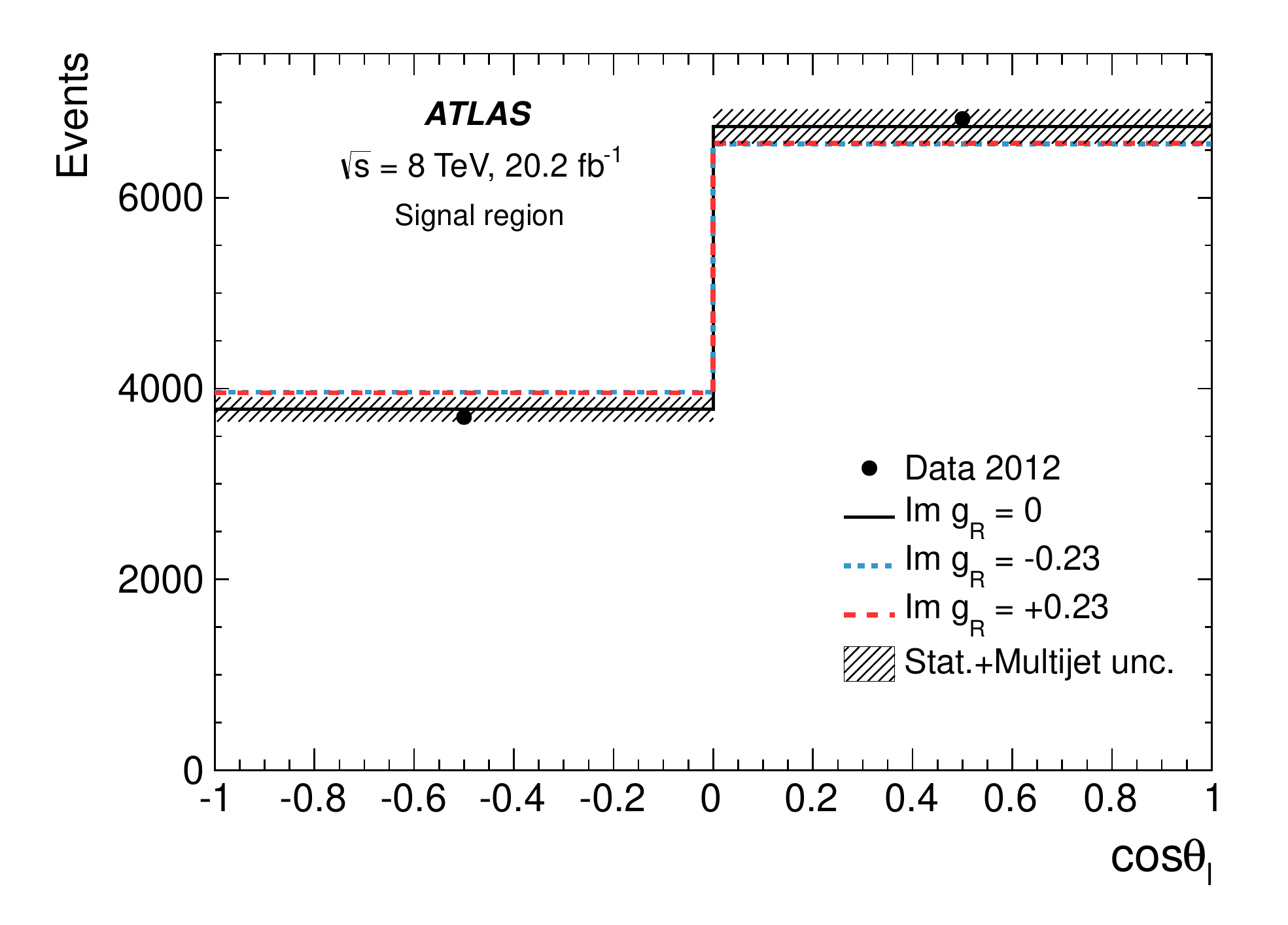}
\label{fig:Protos_cos_lepton_ujet_tchan_1tag_cuts_noFiducial_2jets_Leptons_2}}
\subfloat[]
{\includegraphics[width=0.48\textwidth]{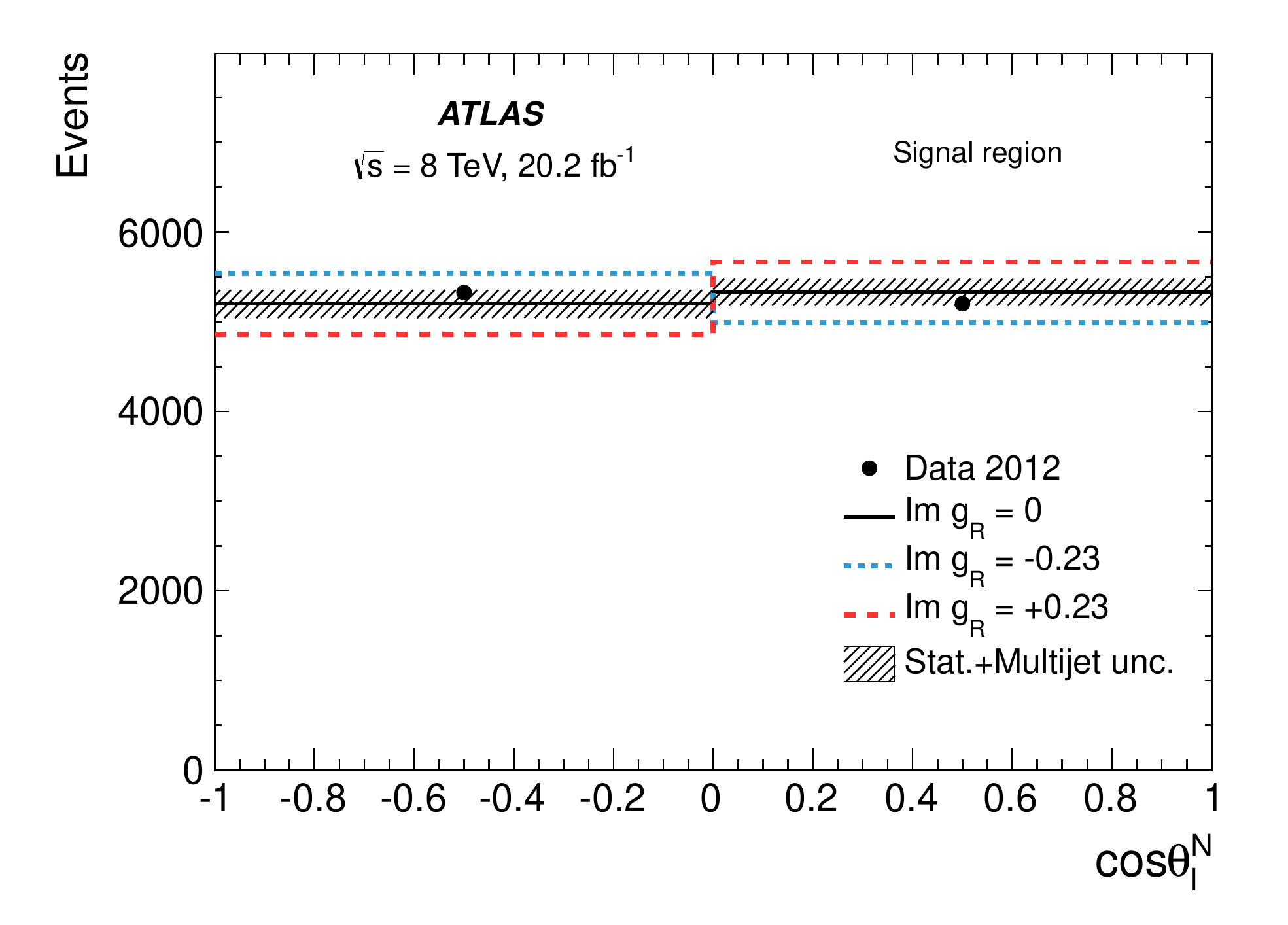}
\label{fig:Protos_cos_lepton_normal_tchan_1tag_cuts_noFiducial_2jets_Leptons_2}}
\caption
{Comparison of the distributions observed in the signal region with the distributions predicted 
as a function of Im\,\gr\ for the angular observables from which the asymmetries used to set 
limits on this coupling are measured:
\protect\subref{fig:Protos_cos_lepton_ujet_tchan_1tag_cuts_noFiducial_2jets_Leptons_2} $\cos\theta_{\ell}$ for $\AFB^{\ell}$ and
\protect\subref{fig:Protos_cos_lepton_normal_tchan_1tag_cuts_noFiducial_2jets_Leptons_2} $\cos\theta_{\ell}^{N}$ for $\AFB^{N}$.
The predicted distributions are determined by adding the signal and background contributions
normalised to the results of the maximum-likelihood fit. The template signal distributions 
are taken from the \textsc{Protos} samples generated with Im\,\gr\ $=0$ (Standard Model 
parameterisation) and Im\,\gr\ $=\pm0.23$. The corresponding parton-level values for the 
$\AFB^{N}$ asymmetry are 0 and $\pm$0.10, respectively. For $\AFB^{\ell}$ the predicted 
values are 0.45 for Im\,\gr\ $=0$ and 0.34 for Im\,\gr\ $=\pm0.23$. The uncertainty bands 
include the statistical post-fit uncertainty, the uncertainty due to the limited size of the 
simulation samples and the uncertainty in the normalisation of the multijet background, added 
in quadrature.}
\label{fig:Distributions_Protos}
\end{figure}

The sensitivity to Im\,\gr\ of the $\cos\theta_{\ell}$ and $\cos\theta_{\ell}^{N}$ 
distributions, which are used to set limits on this coupling, is illustrated in 
\Figure{fig:Distributions_Protos}. In this figure the observed distributions are 
compared to the signal-plus-background predictions built by adding the signal 
templates given by the \textsc{Protos} samples generated with Im\,\gr\ $=0$ (Standard 
Model parameterisation) and Im\,\gr\ $=\pm0.23$, the latter corresponding to the 
maximum values considered in the interpolation method described above.

%% file: Uncertainties.tex
\section{Systematic uncertainties}
\label{sec:Uncertainties}

Several sources of systematic uncertainty affect the asymmetry measurements, modifying
the signal and background event yields and angular distributions. To evaluate the impact 
of each source the asymmetries are extracted by unfolding the template distributions 
after varying them to reflect that source of uncertainty. In each case a new background 
estimation is performed before subtraction, using the fitting procedure described in \Section{sec:Yields}. 
For all sources of systematic uncertainty other than those associated with the limited 
size of the simulation samples, the nominal unfolding corrections are considered. The 
systematic uncertainty is evaluated as the difference between the nominal asymmetry 
value and the one measured using the varied normalisations and shapes.

The sources of systematic uncertainty are split into the following categories:

\textbf{Background normalisation:}
The uncertainties in the normalisation of the top-quark and $W$+jets background processes 
are determined from the maximum-likelihood fit. For the merged $Z$+jets and diboson processes 
the normalisation uncertainty of 20\% introduced in \Section{sec:Backgrounds} is applied to 
the predictions. For the data-driven normalisation of the multijet background the uncertainty 
of 70\% estimated from the comparison of the matrix-method estimates with those given by the 
jet-electron and anti-muon methods is used.

The uncertainty in the integrated luminosity is 1.9\%~\cite{ATLAS_lumi}. It is propagated 
to the asymmetry measurements through the normalisation of the simulated backgrounds.

\textbf{Detector modelling:}
Systematic uncertainties in the reconstruction and energy calibration of jets, electrons 
and muons are propagated in the analysis through variations in the modelling of the detector 
response. For the jets, the main source of uncertainty is the energy scale, evaluated using 
a combination of in situ techniques~\cite{ATLAS_jets_1}. Other jet-related uncertainty sources 
are the modelling of the energy resolution~\cite{ATLAS_jets_2} and reconstruction 
efficiency~\cite{ATLAS_jets_1} (both referred to as jet reconstruction uncertainties), and 
the modelling of the tagging efficiencies of $b$-quark jets, $c$-quark jets and light-flavour 
jets~\cite{ATLAS_btag_2,ATLAS_btag_3}. Uncertainties related to leptons come from trigger, 
identification and isolation efficiencies, as well as from the energy scale and 
resolution~\cite{ATLAS_electrons,ATLAS_muons} (all referred to as lepton reconstruction 
uncertainties). The uncertainties in the energy scale and resolution corrections applied to 
leptons and jets are propagated to the computation of the missing transverse momentum. The 
scale and resolution uncertainties due to soft jets and to contributions of calorimeter 
energy deposits not associated with any reconstructed objects are also considered and 
evaluated independently (they are labelled \MET reconstruction uncertainties). For all 
detector modelling uncertainties, positive and negative uncertainties are estimated 
separately from the corresponding shifts.

\textbf{Signal and background modelling:}
Systematic uncertainties associated with the signal and background modelling are estimated
by comparing event samples from different generators and by varying parameters in the event generation.

The uncertainty in the matrix-element calculation in the simulation of the $t$-channel 
single-top-quark process is estimated by comparing \textsc{MadGraph5}\_a\textsc{MC@NLO}+\textsc{Herwig} 
with \textsc{Powheg-Box}+\textsc{Herwig}. For the \ttbar\ and $Wt$ processes, \textsc{MC@NLO} is 
compared with \textsc{Powheg-Box}, both generators interfaced to \textsc{Herwig}. The uncertainty 
in the parton shower is estimated by comparing \textsc{Powheg-Box} interfaced with \textsc{Pythia} 
and \textsc{Herwig} for the $t$-channel, \ttbar\ and $Wt$ processes. For the $s$-channel 
single-top-quark contribution the uncertainty due to the choice of generator and parton shower is 
estimated in a combined way by comparing \textsc{MC@NLO}+\textsc{Herwig} with 
\textsc{Powheg-Box}+\textsc{Pythia}.

An additional modelling uncertainty is considered for the signal process by comparing the 
NLO {\textsc Powheg-Box} sample to the LO {\textsc Protos} sample implementing the Standard Model 
parameterisation of the $Wtb$ couplings. To estimate this uncertainty, only the shapes of the
distributions are varied in order to assess the impact of using a LO generator to determine 
the unfolding corrections.

The uncertainty in the amount of QCD radiation is evaluated for all top-quark processes by 
comparing the \textsc{Powheg-Box} samples generated with the varied hard-process and parton-shower 
scales presented in \Section{sec:Samples}. The largest shift in the measured asymmetries is 
taken as uncertainty. 

The dependence of the measured asymmetries on the top-quark mass is estimated using \textsc{Powheg-Box} 
samples generated with different top-quark masses. Variations lower than 0.01 per GeV 
are found for the measured asymmetry values. Therefore, these variations are not included 
in the total systematic uncertainty.

The impact of the flavour composition on the modelling of the $W$+jets distributions is 
determined by propagating an uncertainty of 50\% in the ratio of $W$+$b\bar{b}$ to 
$W$+$c\bar{c}$ events. As reported in \Section{sec:Yields}, $W$+light-flavour jets events 
give a small contribution in the signal region and no associated modelling uncertainty 
is taken into account. An additional shape-modelling uncertainty is considered for the 
$W$+jets distributions. Indeed, in the $W$+jets control region a few kinematic variables 
are slightly mismodelled, and the impact of this mismodelling is evaluated by reweighting 
the $W$+jets angular distributions in the signal region. The applied event weights are 
derived from matching to data (after subtraction of all processes other than 
$W$+jets) the mismodelled kinematic variables in the $W$+jets control region. This
procedure leads to a conservative estimate since it also accounts for mismodelling of the 
$W$+light-flavour jets events, which have a much more important contribution in the $W$+jets 
control region than in the signal region. 

The systematic uncertainty associated with the data-driven shape modelling of the multijet events 
is estimated by comparing the shapes provided by the baseline matrix method and the alternative 
modelling given by the jet-electron and anti-muon methods.

All the signal and background modelling uncertainties, except that associated with the
$W$+jets flavour composition, are symmetrised by taking the difference between the nominal 
and varied measurements as positive and negative uncertainties.

Systematic uncertainties related to the parton distribution functions are estimated for all
processes, except for the multijet contribution. The uncertainty is estimated, following the 
PDF4LHC prescription~\cite{PDF4LHC}, by calculating the envelope of the uncertainties at 68\% 
confidence level of the CT10~\cite{PDF_Lai}, MSTW2008NLO~\cite{PDF_Martin} and 
NNPDF2.3~\cite{PDF_Ball} sets.

\textbf{Limited size of simulation samples:}
The uncertainty due to the limited size of the Monte Carlo samples is evaluated by 
varying the background normalisation and shape, as well as the unfolding corrections, 
through Gaussian fluctuations. The standard deviation of the distribution of the 
measured asymmetry provided by an ensemble test of pseudo-experiments built from 
these variations is taken as a systematic uncertainty.

Tables~\ref{tab:Breakdowns_1} and \ref{tab:Breakdowns_2} show the contribution of each 
source of systematic uncertainty to the asymmetry measurements. The total uncertainties 
are obtained from the sum in quadrature of all contributions. Tables~\ref{tab:Breakdowns_1} 
and \ref{tab:Breakdowns_2} also include the statistical uncertainty from the data sample. 
It is evaluated using a procedure similar to that used for the uncertainty associated 
with the size of the simulation samples, but varying the observed numbers of events and 
the shape of the angular distributions through Poisson fluctuations.

The asymmetry measurements are dominated by the systematic uncertainties. The largest 
contributions are from the uncertainties in the modelling of the $t$-channel and 
\ttbar\ processes, and in the jet reconstruction and energy scale. Significant contributions 
also come from the uncertainty in the modelling of the multijet or $W$+jets events,
depending on the measured asymmetry, and from the limited size of the simulation samples. 
The statistical uncertainty of the data sample, although lower than the systematic 
uncertainty, also has a sizeable impact on the measurement precision. 

\begin{table}[!ht]
\begin{center}
\begin{tabular}{lcccc}
\toprule
Uncertainty source                 & $\Delta\AFB^{\ell}\times10^2$ & $\Delta\AFB^{tW}\times10^2$ & $\Delta\AFB\times10^2$ & $\Delta\AEC\times10^2$ \\
\midrule
Statistical uncertainty            & $\pm$2.6      & $\pm$3.1      & $\pm$2.3      & $\pm$2.8 \\
\midrule
Simulation statistics              & $\pm$1.7      & $\pm$1.9      & $\pm$1.4      & $\pm$1.7 \\
Luminosity                         & $<$0.1        & $<$0.1        & $<$0.1        & $<$0.1   \\
Background normalisation           & $\pm$0.5      & $\pm$0.5      & $\pm$0.9      & $\pm$0.7 \\[0.14cm]
\MET reconstruction                & $^{+0.9}_{-0.1}$ & $^{+0.4}_{-0.7}$ & $^{+1.1}_{-0.7}$ & $^{+0.8}_{-0.2}$ \\[0.14cm]
Lepton reconstruction              & $^{+1.0}_{-0.4}$ & $^{+0.1}_{-1.3}$ & $\pm$1.4         & $^{+0.6}_{-0.3}$ \\[0.14cm]
Jet reconstruction                 & $\pm$2.1         & $\pm$2.5         & $\pm$1.2         & $\pm$1.8         \\[0.14cm]
Jet energy scale                   & $^{+1.3}_{-1.2}$ & $^{+2.0}_{-1.6}$ & $^{+3.4}_{-2.7}$ & $^{+2.0}_{-0.7}$ \\[0.14cm]
Jet flavour tagging                & $\pm$0.9      & $\pm$0.3      & $\pm$0.6      & $\pm$0.4 \\
PDF                                & $\pm$0.2      & $<$0.1        & $<$0.1        & $\pm$0.2 \\
$t\bar{t}$ generator               & $\pm$2.3      & $\pm$1.0      & $\pm$0.2      & $\pm$1.2 \\
$t\bar{t}$ parton shower           & $\pm$0.6      & $\pm$0.5      & $\pm$2.7      & $\pm$0.3 \\
$t\bar{t}$ scales                  & $\pm$0.2      & $\pm$0.4      & $\pm$1.2      & $\pm$0.3 \\
$Wt$, $s$-channel generator        & $\pm$1.0      & $\pm$1.1      & $\pm$0.4      & $\pm$0.3 \\
$Wt$, $s$-channel scales           & $\pm$0.9      & $\pm$0.3      & $\pm$0.3      & $\pm$0.3 \\
$t$-channel NLO generator          & $\pm$1.4      & $\pm$0.6      & $\pm$0.6      & $\pm$2.7 \\
$t$-channel LO--NLO generator      & $\pm$1.5      & $\pm$2.0      & $\pm$2.6      & $\pm$1.8 \\
$t$-channel parton shower          & $\pm$0.5      & $\pm$1.0      & $\pm$3.5      & $\pm$0.2 \\
$t$-channel scales                 & $\pm$1.1      & $\pm$2.0      & $\pm$0.6      & $\pm$1.6 \\[0.14cm]
$W$+jets, multijet modelling       & $^{+1.9}_{-2.4}$ & $^{+0.9}_{-1.0}$ & $^{+2.2}_{-2.1}$ & $^{+1.3}_{-1.2}$ \\[0.14cm]
\midrule
Total systematic uncertainty       & $^{+5.4}_{-5.4}$ & $^{+5.2}_{-5.3}$ & $^{+7.3}_{-6.9}$ & $^{+5.3}_{-4.8}$ \\[0.14cm]
\bottomrule
\end{tabular}
\caption{Uncertainties contributing to the measurements of the $\AFB^{\ell}$, $\AFB^{tW}$, $\AFB$ and 
$\AEC$ asymmetries. For better readability the uncertainties are multiplied by 10$^2$.}
\label{tab:Breakdowns_1}
\end{center}
\end{table}

\begin{table}[!ht]
\begin{center}
\begin{tabular}{lcccc}
\toprule
Uncertainty source                 & $\Delta\AFB^{N}\times10^2$ & $\Delta\AFB^{T}\times10^2$ & $\Delta\AFB^{N,\phi}\times10^2$ & $\Delta\AFB^{T,\phi}\times10^2$ \\
\midrule
Statistical uncertainty            & $\pm$2.2      & $\pm$3.1       & $\pm$3.0      & $\pm$4.6 \\
\midrule
Simulation statistics              & $\pm$1.3      & $\pm$2.0       & $\pm$1.8      & $\pm$2.9 \\
Luminosity                         & $<$0.1        & $<$0.1         & $<$0.1        & $<$0.1   \\
Background normalisation           & $\pm$0.4      & $\pm$1.1       & $\pm$0.6      & $\pm$1.1 \\[0.14cm]
\MET reconstruction                & $^{+0.3}_{-0.4}$ & $^{+0.5}_{-0.3}$ & $^{+0.5}_{-0.8}$ & $^{+0.4}_{-1.3}$ \\[0.14cm]
Lepton reconstruction              & $^{+0.1}_{-0.2}$ & $^{+1.3}_{-1.5}$ & $^{+0.6}_{-0.5}$ & $^{+1.6}_{-0.6}$ \\[0.14cm]
Jet reconstruction                 & $\pm$0.8         & $\pm$0.5         & $\pm$1.6         & $\pm$1.3         \\[0.14cm]
Jet energy scale                   & $^{+0.9}_{-0.8}$ & $^{+3.9}_{-4.6}$ & $^{+0.6}_{-2.5}$ & $^{+4.5}_{-2.5}$ \\[0.14cm]
Jet flavour tagging                & $\pm$0.2      & $\pm$0.6       & $\pm$0.3      & $\pm$0.6 \\
PDF                                & $\pm$0.1      & $\pm$0.1       & $\pm$0.1      & $\pm$0.4 \\
$t\bar{t}$ generator               & $\pm$0.2      & $\pm$3.5       & $\pm$1.7      & $\pm$1.3 \\
$t\bar{t}$ parton shower           & $\pm$1.5      & $\pm$1.0       & $\pm$0.9      & $\pm$1.6 \\
$t\bar{t}$ scales                  & $\pm$0.3      & $\pm$0.8       & $\pm$0.3      & $\pm$1.3 \\
$Wt$, $s$-channel generator        & $\pm$0.2      & $\pm$0.8       & $\pm$0.3      & $\pm$1.4 \\
$Wt$, $s$-channel scales           & $\pm$0.6      & $\pm$0.5       & $\pm$0.4      & $\pm$0.9 \\
$t$-channel NLO generator          & $\pm$0.3      & $\pm$4.5       & $\pm$2.6      & $\pm$7.2 \\
$t$-channel LO--NLO generator      & $\pm$0.5      & $\pm$1.9       & $\pm$1.3      & $\pm$3.2 \\
$t$-channel parton shower          & $\pm$0.7      & $\pm$0.9       & $<$0.1        & $\pm$1.1 \\
$t$-channel scales                 & $\pm$0.9      & $\pm$2.2       & $\pm$1.4      & $\pm$2.6 \\[0.14cm]
$W$+jets, multijet modelling       & $^{+0.7}_{-0.6}$ & $^{+1.3}_{-1.7}$ & $\pm$0.6 & $^{+2.3}_{-1.7}$ \\[0.14cm]
\midrule
Total systematic uncertainty       & $^{+2.9}_{-2.9}$ & $^{+8.3}_{-8.8}$ & $^{+4.8}_{-5.4}$ & $^{+10.9}_{-10.1}$ \\[0.14cm]
\bottomrule
\end{tabular}
\caption{Uncertainties contributing to the measurements of the $\AFB^{N}$, $\AFB^{T}$, $\AFB^{N,\phi}$ 
and $\AFB^{T,\phi}$ asymmetries. For better readability the uncertainties are multiplied by 10$^2$.}
\label{tab:Breakdowns_2}
\end{center}
\end{table}

%% file: Results.tex
\section{Results}
\label{sec:Results}

The values of the asymmetries related to the top-quark polarisation and to the 
$W$-boson spin observables, measured using the Standard Model $Wtb$ couplings
for the signal unfolding corrections and for the top-quark background modelling, 
are

\vspace{-0.5cm}
\begin{align*}
\nonumber
\AFB^{\ell}   &=  0.49 \pm 0.03\stat \pm 0.05\syst =  0.49 \pm 0.06\, ,\\
\AFB^{tW}     &=  0.10 \pm 0.03\stat \pm 0.05\syst =  0.10 \pm 0.06\, ,\\
\AFB          &= -0.26 \pm 0.02\stat \pm 0.07\syst = -0.26 \pm 0.08\, ,\\
\AEC          &= -0.25 \pm 0.03\stat \pm 0.05\syst = -0.25 \pm 0.06\, ,\\
\AFB^{T}      &=  0.39 \pm 0.03\stat \pm 0.09\syst =  0.39 \pm 0.09\, ,\\
\AFB^{N,\phi} &= -0.03 \pm 0.03\stat \pm 0.05\syst = -0.03 \pm 0.06\, ,\\
\AFB^{T,\phi} &= -0.17 \pm 0.05\stat ^{+0.11}_{-0.10}\syst = -0.17 ^{+0.12}_{-0.11}\, .
\end{align*}

The values for the top-quark polarisation combined with the charged-lepton spin analysing power 
and with the sum of the $W$-boson helicity fractions, derived from the measured $\AFB^{\ell}$ and 
$\AFB^{tW}$ asymmetries using the relations given in \Table{tab:Asymmetries}, are

\vspace{-0.5cm}
\begin{align*}
\nonumber
\alpha_{\ell} P &= 0.97 \pm 0.05\stat \pm 0.11\syst = 0.97 \pm 0.12\, ,\\
P (\FR+\FL)     &= 0.25 \pm 0.08\stat \pm 0.14\syst = 0.25 \pm 0.16\, .
\end{align*}

The values of the $W$-boson spin observables derived from the measured $\AFB$, $\AEC$, $\AFB^{T}$,
$\AFB^{N,\phi}$ and $\AFB^{T,\phi}$ asymmetries through the relations given in \Table{tab:Asymmetries} 
are

\vspace{-0.5cm}
\begin{align*}
\nonumber
\langle S_3\rangle &= -0.35 \pm 0.03\stat \pm 0.10\syst = -0.35 \pm 0.10\, ,\\
\langle T_0\rangle &= -0.55 \pm 0.06\stat \pm 0.12\syst = -0.55 \pm 0.13\, ,\\
\langle S_1\rangle &=  0.52 \pm 0.04\stat \pm 0.12\syst =  0.52 \pm 0.12\, ,\\
\langle A_2\rangle &= -0.05 \pm 0.05\stat \pm 0.09\syst = -0.05 \pm 0.10\, ,\\
\langle A_1\rangle &=  0.27 \pm 0.07\stat ^{+0.16}_{-0.17}\syst =  0.27 ^{+0.17}_{-0.19}\, .
\end{align*}

The results for the $\AFB^{N}$ asymmetry, which has the highest sensitivity to the anomalous 
$Wtb$ coupling Im\,\gr, and for its associated $W$-boson spin observable are

\vspace{-0.5cm}
\begin{align*}
\nonumber
\AFB^{N}           &= -0.04 \pm 0.02\stat \pm 0.03\syst = -0.04 \pm 0.04\, ,\\
\langle S_2\rangle &=  0.06 \pm 0.03\stat \pm 0.04\syst =  0.06 \pm 0.05\, .
\end{align*}

\noindent
These observables are measured using the signal corrections interpolated with respect to 
Im\,\gr\ as explained in \Section{sec:Unfolding}, and using the Standard Model couplings 
for the top-quark background modelling.

\begin{figure}[t]
\centering
\includegraphics[width=0.50\textwidth]{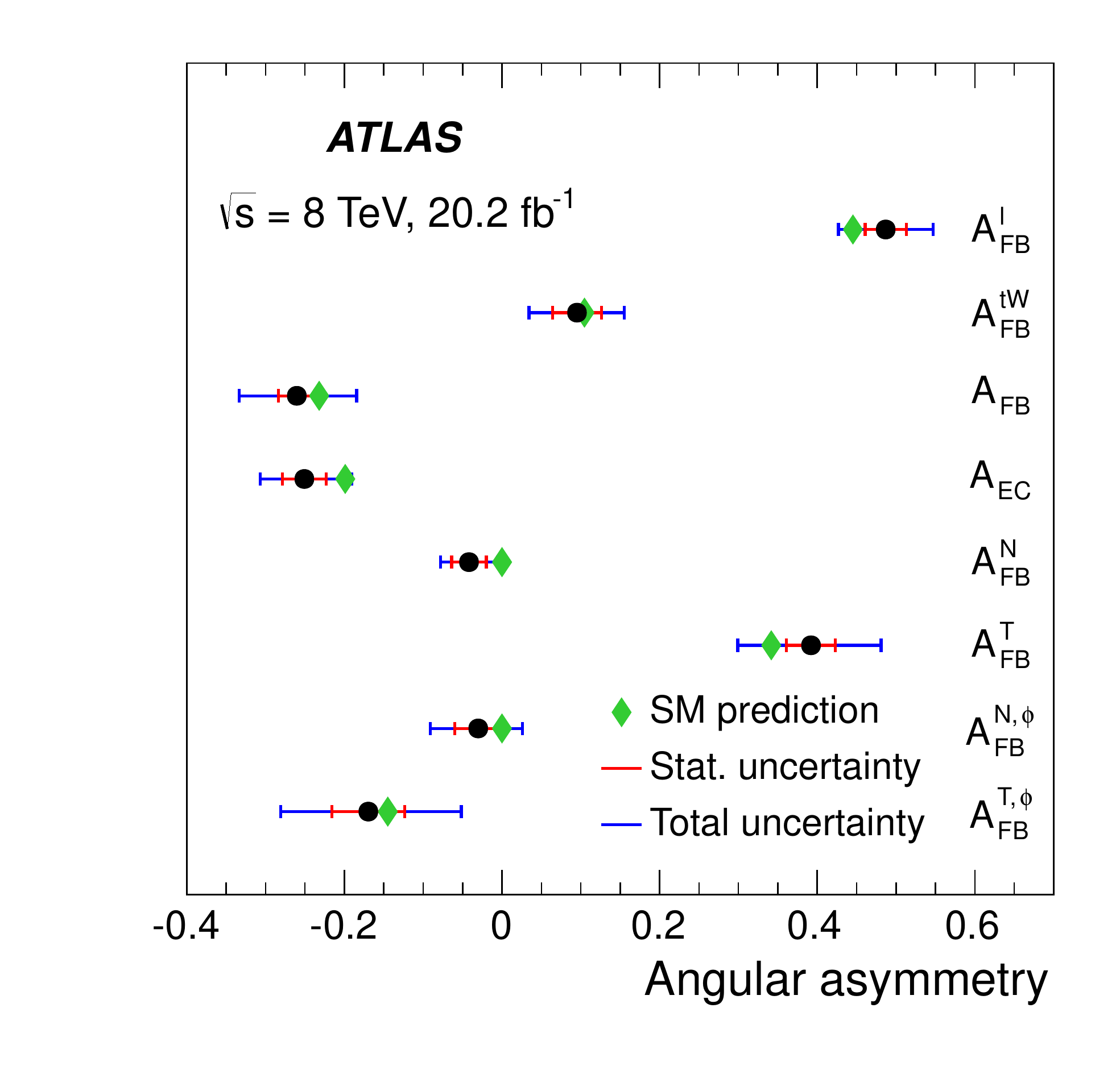}
\caption
{Summary of the measured asymmetries and comparison with the Standard Model predictions.}
\label{fig:Results_asym}
\end{figure}

\begin{figure}[t]
\centering
\includegraphics[width=0.50\textwidth]{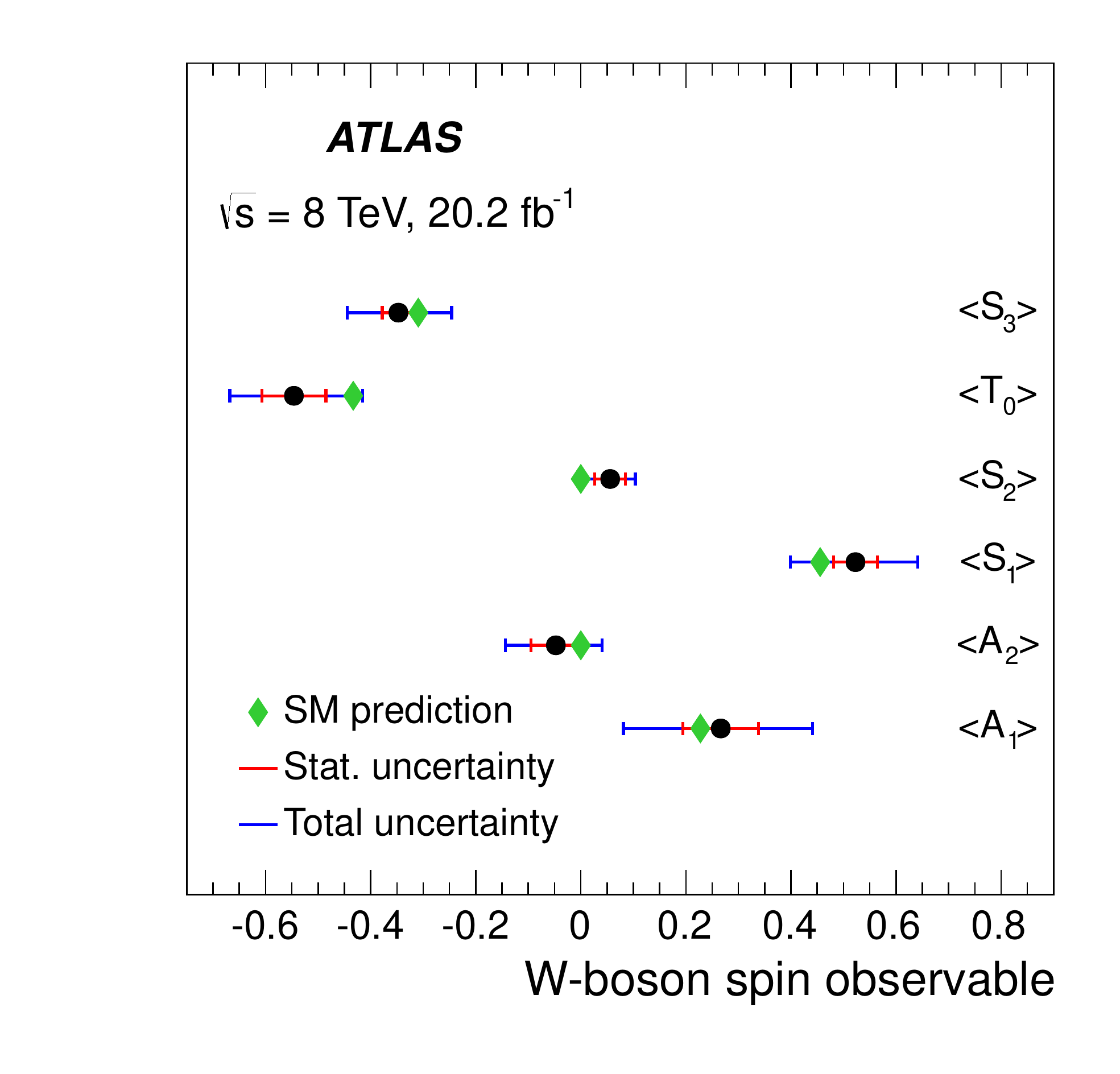}
\caption
{Summary of the measured values of the $W$-boson spin observables and comparison with the Standard 
Model predictions.}
\label{fig:Results_spin}
\end{figure}

\Figure{fig:Results_asym} shows the measured and predicted values of all asymmetries, 
while \Figure{fig:Results_spin} compares the derived values for the six $W$-boson spin 
observables. Compatibility between the measurements and Standard Model predictions is
observed.

The overall compatibility of the measurements with the Standard Model predictions is evaluated 
through the construction of a $\chi^2$ test statistic taking into account all measured quantities 
with their correlations. The theoretical uncertainties, which are negligible compared to the
measurement uncertainties, are not taken into account in the $\chi^2$ calculation. The overall 
covariance matrix is computed from the sum of the covariance matrices associated with the various 
sources of statistical and systematic uncertainty. To calculate the covariance matrices associated 
with the detector-related and $W$+jets flavour composition uncertainties, the positive and 
negative uncertainties are symmetrised by taking the larger value. The overall $p$-value for the 
eight asymmetries is found to be 0.94, and it is 0.83 for the six $W$-boson spin observables.

Limits on the anomalous coupling Im\,\gr\ are extracted from the $\AFB^{N}$ and $\AFB^{\ell}$ 
asymmetries, which, as discussed in \Section{sec:Unfolding}, are measured independently of any 
assumption about Im\,\gr\ in the unfolding procedure, but assuming the Standard Model couplings
for the subtracted top-quark backgrounds. However, for the main \ttbar\ background a negligible 
dependence on Im\,\gr\ is expected.

The limit extraction is based on the \textsc{TopFit} code~\cite{Theory_Wpolarization,Comb_Wtb_limits}. 
By taking into account the analytic expressions and parameterisations introduced in 
Refs.~\cite{Theory_Wpolarization,Theory_Wspin,Theory_TopSpin_2} for the $Wtb$ coupling dependence 
of $\langle S_2\rangle$, $\alpha_{\ell}$ and $P$, it is possible to determine the allowed region 
for Im\,\gr\ from the measured values of $\AFB^{N}$ and $\AFB^{\ell}$. The limit setting is based
on the computation of the $\chi^2$ test statistic using the covariance matrix associated with the 
$\AFB^{N}$ and $\AFB^{\ell}$ measurements. An overall correlation coefficient of $-$0.05 is found.

Assuming $\vl=1$ and that all anomalous couplings other than Im\,\gr\ vanish ($\vr=\gl=0$ and 
Re\,$\gr=0$), the limits set at the 95\% confidence level are Im\,\gr~$\in$~[$-$0.18, 0.06].
The measured interval of allowed values slightly improves on the limits set at 7~TeV by the ATLAS 
Collaboration from the measurement of double-differential angular decay rates~\cite{ATLAS_Wtb_limits}.

%% file: Conclusion.tex
\section{Conclusion}
\label{sec:Conclusion}

Measurements of the top-quark and $W$-boson polarisation observables in $t$-channel single 
top-quark production at $\sqrt{s}=8$~TeV with 20.2~fb$^{-1}$ of $pp$ collision data recorded 
with the ATLAS detector at the LHC are presented. The selected events contain one isolated 
electron or muon, large missing transverse momentum and exactly two jets, of which one is 
tagged as a $b$-jet. A cut-based analysis is used to discriminate 
the signal events from background, and the electron and muon channels are combined. The 
polarisation observables are measured from asymmetries in various angular distributions 
unfolded to the parton level. Unfolding corrections based on a Standard Model simulation of the 
$t$-channel process are used, as well as model-independent corrections derived through an 
interpolation method. The measured asymmetries and the measured polarisation observables are 
in agreement with the predictions of the Standard Model. Limits on the imaginary part of 
the anomalous coupling \gr\ are also set, giving Im\,\gr~$\in$~[$-$0.18, 0.06] at the 95\% 
confidence level. The extracted values improve on the most recently published limits for this
coupling.

%% file: Acknowledgements.tex

We thank CERN for the very successful operation of the LHC, as well as the
support staff from our institutions without whom ATLAS could not be
operated efficiently.

We acknowledge the support of ANPCyT, Argentina; YerPhI, Armenia; ARC, Australia; BMWFW and FWF, Austria; ANAS, Azerbaijan; SSTC, Belarus; CNPq and FAPESP, Brazil; NSERC, NRC and CFI, Canada; CERN; CONICYT, Chile; CAS, MOST and NSFC, China; COLCIENCIAS, Colombia; MSMT CR, MPO CR and VSC CR, Czech Republic; DNRF and DNSRC, Denmark; IN2P3-CNRS, CEA-DSM/IRFU, France; GNSF, Georgia; BMBF, HGF, and MPG, Germany; GSRT, Greece; RGC, Hong Kong SAR, China; ISF, I-CORE and Benoziyo Center, Israel; INFN, Italy; MEXT and JSPS, Japan; CNRST, Morocco; FOM and NWO, Netherlands; RCN, Norway; MNiSW and NCN, Poland; FCT, Portugal; MNE/IFA, Romania; MES of Russia and NRC KI, Russian Federation; JINR; MESTD, Serbia; MSSR, Slovakia; ARRS and MIZ\v{S}, Slovenia; DST/NRF, South Africa; MINECO, Spain; SRC and Wallenberg Foundation, Sweden; SERI, SNSF and Cantons of Bern and Geneva, Switzerland; MOST, Taiwan; TAEK, Turkey; STFC, United Kingdom; DOE and NSF, United States of America. In addition, individual groups and members have received support from BCKDF, the Canada Council, CANARIE, CRC, Compute Canada, FQRNT, and the Ontario Innovation Trust, Canada; EPLANET, ERC, ERDF, FP7, Horizon 2020 and Marie Sk{\l}odowska-Curie Actions, European Union; Investissements d'Avenir Labex and Idex, ANR, R{\'e}gion Auvergne and Fondation Partager le Savoir, France; DFG and AvH Foundation, Germany; Herakleitos, Thales and Aristeia programmes co-financed by EU-ESF and the Greek NSRF; BSF, GIF and Minerva, Israel; BRF, Norway; CERCA Programme Generalitat de Catalunya, Generalitat Valenciana, Spain; the Royal Society and Leverhulme Trust, United Kingdom.

The crucial computing support from all WLCG partners is acknowledged gratefully, in particular from CERN, the ATLAS Tier-1 facilities at TRIUMF (Canada), NDGF (Denmark, Norway, Sweden), CC-IN2P3 (France), KIT/GridKA (Germany), INFN-CNAF (Italy), NL-T1 (Netherlands), PIC (Spain), ASGC (Taiwan), RAL (UK) and BNL (USA), the Tier-2 facilities worldwide and large non-WLCG resource providers. Major contributors of computing resources are listed in Ref.~\cite{ATL-GEN-PUB-2016-002}.

%% file: atlas_authlist.tex
\begin{flushleft}
{\Large The ATLAS Collaboration}

\bigskip

M.~Aaboud$^\textrm{\scriptsize 137d}$,
G.~Aad$^\textrm{\scriptsize 88}$,
B.~Abbott$^\textrm{\scriptsize 115}$,
J.~Abdallah$^\textrm{\scriptsize 8}$,
O.~Abdinov$^\textrm{\scriptsize 12}$,
B.~Abeloos$^\textrm{\scriptsize 119}$,
O.S.~AbouZeid$^\textrm{\scriptsize 139}$,
N.L.~Abraham$^\textrm{\scriptsize 151}$,
H.~Abramowicz$^\textrm{\scriptsize 155}$,
H.~Abreu$^\textrm{\scriptsize 154}$
% ...
\bigskip
\\
% ...
$^{8}$ Department of Physics, The University of Texas at Arlington, Arlington TX, United States of America\\
% ...
$^{12}$ Institute of Physics, Azerbaijan Academy of Sciences, Baku, Azerbaijan\\
% ...
$^{88}$ CPPM, Aix-Marseille Universit{\'e} and CNRS/IN2P3, Marseille, France\\
% ...
$^{115}$ Homer L. Dodge Department of Physics and Astronomy, University of Oklahoma, Norman OK, United States of America\\
% ...
$^{119}$ LAL, Univ. Paris-Sud, CNRS/IN2P3, Universit{\'e} Paris-Saclay, Orsay, France\\
% ...
$^{137}$ $^{(a)}$ Facult{\'e} des Sciences Ain Chock, R{\'e}seau Universitaire de Physique des Hautes Energies - Universit{\'e} Hassan II, Casablanca; $^{(b)}$ Centre National de l'Energie des Sciences Techniques Nucleaires, Rabat; $^{(c)}$ Facult{\'e} des Sciences Semlalia, Universit{\'e} Cadi Ayyad, LPHEA-Marrakech; $^{(d)}$ Facult{\'e} des Sciences, Universit{\'e} Mohamed Premier and LPTPM, Oujda; $^{(e)}$ Facult{\'e} des sciences, Universit{\'e} Mohammed V, Rabat, Morocco\\
% ...
$^{139}$ Santa Cruz Institute for Particle Physics, University of California Santa Cruz, Santa Cruz CA, United States of America\\
% ...
$^{151}$ Department of Physics and Astronomy, University of Sussex, Brighton, United Kingdom\\
% ...
$^{154}$ Department of Physics, Technion: Israel Institute of Technology, Haifa, Israel\\
$^{155}$ Raymond and Beverly Sackler School of Physics and Astronomy, Tel Aviv University, Tel Aviv, Israel\\
% ...
$^{a}$ Also at Department of Physics, King's College London, London, United Kingdom\\
$^{b}$ Also at Institute of Physics, Azerbaijan Academy of Sciences, Baku, Azerbaijan\\
$^{c}$ Also at Novosibirsk State University, Novosibirsk, Russia\\
$^{d}$ Also at TRIUMF, Vancouver BC, Canada\\
$^{e}$ Also at Department of Physics {\&} Astronomy, University of Louisville, Louisville, KY, United States of America\\
% ...
\end{flushleft}
